\documentclass[11pt]{article}
\pdfoutput=1
\usepackage{jcappub,natbib}
\input{colordvi.tex}

\def\ga{\mathrel{\raise.3ex\hbox{$>$\kern-.75em\lower1ex\hbox{$\sim$}}}}
\def\la{\mathrel{\raise.3ex\hbox{$<$\kern-.75em\lower1ex\hbox{$\sim$}}}}

\newcommand{\lp}{\left (}
\newcommand{\rp}{\right )}
\newcommand{\llp}{\left [}
\newcommand{\rrp}{\right ]}
\newcommand{\I}{{\cal I}}
\newcommand{\Ic}{{\cal I}_c}
\newcommand{\Is}{{\cal I}_s}
\newcommand{\Pz}{{\cal P}_\zeta}

\def\NL{\text{\tiny NL}}

\def\PBH{\text{\tiny PBH}}
\def\GW{\text{\tiny GW}}
\renewcommand{\la}{\left \langle}
\newcommand{\ra}{\right \rangle}

\usepackage{hyperref}
\newcommand{\arXiv}[2]{\href{http://arxiv.org/pdf/#1}{{\tt [#2/#1]}}}
\newcommand{\arXivold}[1]{\href{http://arxiv.org/pdf/#1}{{\tt [#1]}}}

\title{Title}

\author[a,b,c]{N. Bartolo,}
\author[a]{D. Bertacca,}
\author[d]{V. De Luca,}
\author[d]{G. Franciolini,}
\author[a,b,c,e]{S. Matarrese,}
\author[a,b]{M. Peloso,}
\author[b]{A. Ricciardone,}
\author[d,f]{A. Riotto,} 
\author[g]{G.~Tasinato}

\affiliation[a]{Dipartimento di Fisica e Astronomia “Galileo Galilei” Universita` di Padova, 35131 Padova, Italy}
\affiliation[b]{INFN, Sezione di Padova, 35131 Padova, Italy}
\affiliation[c]{INAF - Osservatorio Astronomico di Padova, I-35122 Padova, Italy}
\affiliation[d]{Department of Theoretical Physics and Center for Astroparticle Physics (CAP), CH-1211 Geneva 4, Switzerland}
\affiliation[e]{Gran Sasso Science Institute, I-67100 L'Aquila, Italy}
\affiliation[f]{CERN,Theoretical Physics Department, Geneva, Switzerland}
\affiliation[g]{Department of Physics, Swansea University, Swansea, SA2 8PP, UK}

\title{Gravitational Wave Anisotropies from Primordial Black Holes}

\abstract{An observable stochastic background of gravitational waves is generated whenever primordial black holes are created in the early universe thanks to a small-scale enhancement of the curvature perturbation. We calculate the anisotropies and  non-Gaussianity of such stochastic gravitational waves background which receive two contributions, the first at formation time and the second due to propagation effects. The former contribution can be generated if the distribution of the curvature perturbation is characterized by  a local and  scale-invariant 
  shape of non-Gaussianity.
Under such an assumption, we conclude that  a sizeable magnitude of anisotropy and non-Gaussianity in the gravitational waves would suggest that primordial black holes may not comply the totality of the dark matter.}

\begin{document}

\maketitle
\flushbottom

\section{Introduction}
\label{sec:intro} 

Following the first measurements of the gravitational waves (GWs) generated by $\sim {\cal O } \left( 10 \right) M_\odot$ black-hole mergers \cite{Abbott:2016blz}, the past few years have  witnessed a renewed interest in Primordial Black Holes (PBHs) \cite{Bird:2016dcv,Garcia-Bellido:2017fdg,Sasaki:2018dmp,Barack:2018yly}. Bounds of various origins exist on the PBHs abundance for a wide range of PBHs masses \cite{Sasaki:2018dmp}, leaving also open the possibility that PBHs in certain mass ranges could be identified with a substantial fraction or, possibly, the totality of the dark matter of the universe. This is particularly true for PBH of masses of  $\sim {\cal O } \left( 10^{-12} \right) M_\odot$, for which previously expected limits from femtolensing  \cite{Katz:2018zrn} and dynamical constraints from White Dwarves \cite{Montero-Camacho:2019jte}
have been shown to be invalid.

A simple mechanism for PBHs generation is from enhanced density perturbations $\delta \rho$ produced during inflation. If, using standard arguments from (nearly) scale invariance, we extrapolate the power of the perturbations $P_\zeta \sim \left( \delta \rho/\rho \right)^2 = {\cal O } \left( 10^{-9} \right)$ measured at CMB scales to small scales, we obtain a completely negligible fraction of PBHs. On the other hand, an increase of this power can strongly increase the portion of the universe which, at horizon re-entry, have an energy density above the threshold that leads to the collapse and PBH formation \cite{th1,th2,th3,th4}. This increase would be associated to a breaking of scale invariance at some given scale, which in turn reflects some specific dynamical mechanism taking place during inflation  \cite{s1,s2,s3,Espinosa:2017sgp} (see \cite{Sasaki:2018dmp} for a review and references therein). 

The increased density perturbations unavoidably lead to GWs production due to the intrinsic nonlinear nature of gravity 
\cite{Acquaviva:2002ud,Mollerach:2003nq,Ananda:2006af,Baumann:2007zm,e,Saito:2009jt,altri8}~\footnote{It is important to stress that we do not refer here to the GWs produced only in the regions that collapse to form the PBH, but from everywhere in the universe, due to the general increased of the power of the density perturbations \cite{DeLuca:2019llr}.} . This GW emission can be used to constrain the PBH abundance  \cite{Saito:2009jt}. In fact, let us assume that the power $P_\zeta$ of the primordial scalar perturbations has an enhancement at some give scale $k_*$, leading to a significant fraction $f_{\PBH}$ of dark matter and also to an observable amount of GWs. Then, even a small decrease of $P_\zeta$ from this level would lead to a completely negligible value for $f_{\PBH}$~\footnote{This is due to the fact that, assuming Gaussian primordial perturbations, only the rare regions with $\delta \rho \gg  \sigma$, being $\sigma$ the square root of the variance, have an energy density above the threshold for PBH formation. A change of the variance have a strong impact on the area of the tail of the distribution above the PBH threshold.  For an example of a case with non-Gaussian primordial perturbations, see, e.g.,~\cite{Garcia-Bellido:2017aan}.} with a very minor change of the amount of GWs. Therefore GW observations are sensitive even to peaks in $P_\zeta$ that are associated to a very small (possibly, otherwise unobservable) amount of PBHs. 
The characteristic frequency of the GWs emitted by the production of PBH of mass $M$ is $f\simeq 3 \times 10^{-9}\, {\rm Hz} \,(M/M_\odot)^{-1/2}$ \cite{Saito:2009jt}.

Given the potential relevance of these observations in upcoming experiments like LISA \cite{Audley:2017drz} and DECIGO \cite{Kawamura:2006up}, it is important to characterize the stochastic background of gravitational waves (SGWB) produced with this mechanism \cite{Caprini:2019pxz,Bartolo:2018qqn,noiLISA1,noiLISA2} \footnote{Other mechanisms to generate a SGWB from the early universe can be found in Refs.~\cite{altri1,altri2,altri3,altri4,altri5,altri6,altri7,altri8}.}.
Is it homogeneous in space? Does its spatial distribution obey a Gaussian statistics? To our knowledge, these questions have not yet been addressed for the  SGWB studied in this paper. This is the purpose of this work. 

Even assuming a completely homogeneous and isotropic SGWB at its production, these GWs propagate in a perturbed universe. As a consequence, the GW signal arriving to Earth has angular anisotropies \cite{Alba:2015cms,Contaldi:2016koz,Bertacca:2017vod,Cusin:2017fwz,Jenkins:2018lvb,Cusin:2018avf,Renzini:2019vmt} which are non-Gaussian \cite{Bartolo:2019oiq}. In addition, as we show and quantify in this work, the GW production itself has some degree of anisotropy and non-Gaussianity. A necessary condition for large scale anisotropies and non-Gaussianity is the presence of large-scale perturbations that are needed to produce correlations on cosmological scales, much greater than the scale $k_*^{-1}$ associated to the typical regions forming the PBHs. 
The GW formation is a local event, that, by the equivalence principle, cannot be locally affected by modes of wavelength much greater than the PBH horizon. However, non-Gaussianity of the primordial density perturbations can lead to small-long scale correlations, so that long modes can lead to a large-scale modulation of the local power of the density perturbations and, consequently, on the amount of GWs produced within each region. 

We show here that an amount of (local) non-Gaussianity of the scalar perturbations compatible with the current upper bounds from Planck  \cite{Akrami:2019izv} can lead to an amount of anisotropies and non-Gaussianity of the GWs distribution greater than that due to the propagation  \cite{Bartolo:2019oiq}. On the other hand, if the PBHs constitute a significant fraction of the dark matter, additional limits from isocurvature apply, leading to much stronger limits on the scalar non-Gaussianity \cite{Young:2015kda}. This significantly limits the SGWB anisotropy and non-Gaussianity imprinted at the SWGB production. Therefore, our prediction is that a significant amount of PBH dark matter is associated with a SGWB that is isotropic and Gaussian, up to propagation effects. A stronger amount of anisotropy and non-Gaussianity of the SGWB would signify the existence of a local enhancement of the density perturbations, and of a PBH population that is well below the dark matter abundance. These conclusions   hold  under the strict assumption 
  of
  local, scale-invariant primordial non-Gaussianity in the curvature perturbations, extending from CMB scales down to the small scales relevant for PBH formation. On the other hand, given the huge range of scales involved, different conditions for structure formation might hold, especially 
on the smallest scales, that might break the assumption of scale-invariant non-Gaussianity. This would leave open the possibility of relaxing our constraints, and to allow for PBHs to be the totality of the observed dark matter, with an accompanying SGWB that might still be anisotropic and non-Gaussian.

The paper is organized as follows. In Section \ref{sec:GW2nd}  we review the mechanism of GWs production at second order from scalar density perturbations. In Section \ref{sec:fNL} we compute the amount of anisotropy and non-Gaussianity of the SGWB produced by this mechanism, in the case in which the primordial density perturbations are non-Gaussian. In Section \ref{sec:iso} we review the additional limits on the scalar non-Gaussianity that are present if the PBHs constitute a significant portion of the dark matter. In Section \ref{sec:results} we present a summary of our results and of the existing constraints. Finally, in Section \ref{sec:conclusions} we provide some final remarks. The paper is concluded by two appendices where we present some technical steps of our computations.

\section{GWs at second-order from enhanced density perturbations}
\label{sec:GW2nd} 

A simple mechanism for the production of a distribution of PBHs peaked at a given mass is to assume that some inflationary mechanism has produced a peak of the primordial density perturbations at some given scale. This enhancement reflects some specific dynamical mechanism that took place at some given moment during inflation, thus breaking the approximate scale invariance for modes that exited the horizon at that specific moment. This enhancement increases the amount of regions where, at horizon re-entry of this mode
(in the radiation dominated era, well after inflation) the energy density is above the necessary threshold to produce PBHs, thus increasing the PBH density. We introduce the  power spectrum  for the primordial density perturbations as
\begin{equation}
\left\langle \zeta( \vec{k})  \zeta ( \vec{k}') \right\rangle = \frac{2 \pi^2}{k^3} \, {\cal P}_\zeta \left( k \right) \left( 2 \pi \right)^3 \delta^{(3)}( \vec{k} + \vec{k}' ) \;\;\;,\;\;\; 
{\cal P}_\zeta \left( k \right) = {\cal P}_{\zeta_s} \left( k \right) + {\cal P}_{\zeta_L} \left( k \right),
\label{calP}
\end{equation} 
where $P_{\zeta_L} \left( k \right)$ is the power spectrum of the standard (nearly) scale-invariant perturbations generated during inflation. The suffix ``$L$'' in this term stands for long-wavelength modes, relevant at cosmological scales, which are much greater than the scale $k_*^{-1}$ of the modes forming the PBH, which are labelled with the suffix ``$s$'' and are related to the small-scale power spectrum $P_{\zeta_s} \left( k \right)$. At the short scale $k_*^{-1}$ these long modes are completely subdominant with respect to those contributing to the first term in (\ref{calP}). So they play no role in the local PBH formation and in the local production of GWs that we discuss next. However, as we see in the next Section, in presence of primordial non-Gaussianity these modes can add a long-scale modulation to this production, thus resulting in anisotropies of the SGWB. 

The necessity of local non-Gaussianity to create anisotropies of the SGWB is crucial due to the generation of a cross-talk between the short scale $k_*^{-1}$, of the order of the horizon scale at PBH production,  and the long wavelength scale  $q^{-1}$,  associated to $\zeta_L$. If absent, 
the long scalar modes of wavelength of cosmological size do not change  the local physics in each patch of size  $k_*^{-1}$, and so the amount of PBHs and the induced   GWs is locally the same in any patch. This is simply due to the Equivalence Principle which also dictates that the anisotropies in the   SGWB should decay like $(q/k_*)^2$. 

To have a confirmation of such a general result we present, in the following,  the calculation of the contribution from the 
 enhanced scalar modes in (\ref{calP})  to the production of GWs at second-order, without primordial non-Gaussianity.  This leads to the GW energy density operator
\cite{Ando:2017veq,Espinosa:2018eve} \footnote{One additional contribution to the GWs abundance, which is not considered in this paper, is related to the contraction of peaks in the density fluid, generated by the same curvature perturbations which are responsible for the production of GWs at second-order, which are not high enough to collapse into PBHs (see Ref. \cite{DeLuca:2019llr} for details).}
\begin{align}
&\rho_\GW \left( \eta ,\, \vec{x} \right)  =   
  \frac{ M_p^2}{81  \eta^2 a^2}  \int \frac{d^3 k_1 d^3 k_2 d^3 p_1 d^3 p_2}{\left( 2 \pi \right)^{12} }
\frac{1}{k_1^2 k_2^2} 
 \, {\rm e}^{i \vec{x} \cdot \left( \vec{k}_1 + \vec{k}_2 \right)} 
 T[ {\hat k}_1 ,\, {\hat k}_2 ,\, \vec{p}_1 ,\, \vec{p}_2]
\nonumber\\ 
&
\times
 \zeta( \vec{p}_1)  \zeta ( \vec{k}_1 - \vec{p}_1) 
\zeta( \vec{p}_2)  \zeta( \vec{k}_2 - \vec{p}_2)
\Big<
\prod_{i=1}^2
\left[ {\cal I}_s ( \vec{k}_i ,\, \vec{p}_i) \cos ( k_i \eta) -  {\cal I}_c( \vec{k}_i ,\, \vec{p}_i) \sin( k_i \eta) \right] \Big>_T .
\label{rhox}
\end{align} 
This expression is valid during radiation domination, and the function $T \left[ {\hat k}_1 ,\, {\hat k}_2 ,\, \vec{p}_1 ,\, \vec{p}_2 \right]$ is obtained from a contraction between the internal momenta and the GW polarization operators (we provide the expression in Appendix \ref{app:conventions}, where we also outline our conventions. In that Appendix, we also provide the analytic expressions for 
${\cal I}_{c,s}$ \cite{Espinosa:2018eve,Kohri:2018awv}). The angular brackets in the second line denote a time average, that is necessary for the definition of the energy density in GW \cite{Misner:1974qy,Maggiore:1999vm,Flanagan:2005yc}, and it is performed on a timescale $T$ much greater than the GW phase oscillations ($T k_i \gg 1$) but much smaller than the cosmological time ($T H \ll 1$). Finally, we note the presence of four density perturbation operators $\zeta$ in this expression. This is due to the fact that the GW energy density is a bilinear in the GW field (see Eq. (\ref{rho}), and the GW field sourced at second-order is a bilinear in $\zeta$ (see Eq. (\ref{h-sourced})). 

The one-point expectation value of the operator (\ref{rhox}) is the expected GW energy density from this mechanism. As already mentioned, we assume that the scalar perturbations $\zeta$ are Gaussian (this assumption will be relaxed in the next section) so that the four point function $\left\langle \zeta^4 \right\rangle$ emerging from the expectation value of (\ref{rhox}) can be written as sum of three terms, each containing two products $\left\langle \zeta^2 \right\rangle$. Schematically, 
\begin{equation}
\label{4zeta}
{\rm Gaussian \; } \zeta \;\;\; \Rightarrow \;\;\; 
\left\langle \zeta^4 \right\rangle = 
\left\langle \zeta^2 \right\rangle \, \left\langle \zeta^2 \right\rangle + 
\left\langle \zeta^2 \right\rangle \, \left\langle \zeta^2 \right\rangle + 
\left\langle \zeta^2 \right\rangle \, \left\langle \zeta^2 \right\rangle,
\end{equation}
with all possible permutations of the four operators. One contraction gives a vanishing contribution at finite momentum, while the other two contractions give an identical contribution, and (using the definition of the power spectrum in (\ref{calP})) lead to 
\begin{align} 
&\left\langle \rho_\GW \left( \eta ,\, \vec{x} \right) \right\rangle
\equiv \rho_c ( \eta) \, \int d \ln k \; \Omega_\GW \left( \eta ,\, k \right) \nonumber\\
&=  
  \frac{2 \pi^4 M_p^2}{81  \eta^2 a^2}  \, \int \frac{d^3 k_1 d^3 p_1}{\left( 2 \pi \right)^{6} } 
\frac{1}{k_1^4}\, 
\frac{\left[ p_1^2 -  ( \vec{k}_1 \cdot \vec{p}_1)^2/k_1^2 \right]^2}{p_1^3 \, \left\vert \vec{k_1} - \vec{p}_1 \right\vert^3} \, 
 {\cal P}_\zeta ( p_1) 
{\cal P}_\zeta( |\vec{k_1} - \vec{p}_1|)   \llp{\cal I}_c^2( \vec{k}_1 ,\, \vec{p}_1) + {\cal I}_s^2( \vec{k}_1 ,\, \vec{p}_1) \rrp 
. 
\label{rho1-par}
\end{align} 
The contraction forced $k_1 = k_2$  in Eq. (\ref{rhox}) using Eq.~\eqref{calP}. In this case the time average procedure in (\ref{rhox}) became straightforward, namely 
\begin{equation}
\left\langle \sin^2 \left( k_1 \, \eta \right) \right\rangle_T = \left\langle \cos^2 \left( k_1 \, \eta \right) \right\rangle_T = 1/2 \;,\;\; 
\left\langle \sin \left( k_1 \, \eta \right) \cos \left( k_1 \, \eta \right)  \right\rangle_T = 0. 
\end{equation}
Following the  standard convention, in the first line of (\ref{rho1-par}) we defined the fractional energy density in the GW for log interval. The quantity $\rho_c = 3 H^2 M_p^2$ denotes the critical energy density of a spatially flat universe with Hubble rate $H$. The notation in the second line of (\ref{rho1-par}) exploits the fact that the integral over the two angles $d \Omega_{k}$ can be made trivial by exploiting that the only angular dependence of the integrand is on the angle between $\vec{k}_1$ and $\vec{p}_1$ (this is a consequence of the statistical isotropy of the background). By introducing the rescaled magnitudes $x \equiv p_1 / k_1$ and  $y \equiv |\vec{k}_1 - \vec{p}_1| / k_1$, the expression (\ref{rho1-par}) reduces to \cite{Espinosa:2018eve} 
\begin{equation}
\Omega_{\rm  GW} (k,\eta) = \frac{1}{972a^2 H^2 \eta^2 } 
 \iint _{\cal S} dx dy   \frac{x^2}{ y^2 } \llp  1- \frac{\lp 1+ x^2 - y^2 \rp ^2}{4 x^2}\rrp ^2
{\cal P}_\zeta \left( k x \right) 
{\cal P}_\zeta \left( k y  \right) {\cal I}^2 \left(x,y \right) ,
\end{equation}
where the integration region  ${\cal S}$ extends to $x > 0$ and to $\left\vert 1 - x \right\vert \leq y \leq 1 + x$ and where we defined $\I^2 \equiv \I_c^2 + \I_s^2$. For a Dirac delta 
power spectrum of the scalar curvature perturbation on small scales, ${\cal P}_{\zeta_s} \left( k \right) = A_s \,k_* \, \delta \left( k - k_* \right)$, this expression then becomes~\footnote{This expression is valid during radiation domination; we see that it is costant, and independent of the normalization of the scale factor. } 
\begin{eqnarray} \label{monopole}
\Omega_{\GW} (k,\eta) &=& \frac{1}{a^2 H^2 \eta^2 } \frac{A_s^2}{15552} \, 
  \frac{k^2}{k_*^2} \llp   \frac{4 k_*^2}{k^2}-1  \rrp ^2
  \theta \left( 2 k_* - k \right) \; \I^2\left ( \frac{k_*}{k} , \frac{k_*}{k}\right) \; 
 \end{eqnarray} 
where $\theta$ is the Heaviside step function, and 
\begin{align}
&\I^2\left ( \frac{k_*}{k} , \frac{k_*}{k}\right) 
 \equiv \I_c^2\left ( \frac{k_*}{k} , \frac{k_*}{k}\right) + \I_s^2\left ( \frac{k_*}{k} , \frac{k_*}{k}\right)
\nonumber \\
&  =
\frac{729 }{16} \lp\frac{ k}{ k_*} \rp^{12}
\lp
3-\frac{2 k_*^2}{k^2}\right)^4
 \left\{ \left[ 
4 \left(2 -3 \frac{k^2}{k_*^2}\right)^{-1} -\log \left( \left|   1-\frac{4 k_*^2}{3 k^2}\right| \right)\right]^2+\pi ^2 \theta
   \left(\frac{2 k_*}{\sqrt{3} k}-1\right)
   \right \}. 
\end{align}
  We note that the result (\ref{monopole}) for the one-point expectation value of the GW energy density is independent of position. This follows from statistical homogeneity of the FLRW background universe (at the technical level, it is due to the fact that the contraction of the four $\zeta$ operators in Eq. (\ref{rhox}) forces $\vec{k}_1 + \vec{k}_2=0$). However, as explained in the introduction, one does not expect that the sourced GWs are perfectly homogeneous across the universe. As a consequence, the SGWB reaching us from different directions will present some angular anisotropies. 

To quantify the level of these anisotropies one needs to compute the two-point correlation function $\left\langle \rho_\GW \left( \vec{x} \right)  \rho_\GW \left( \vec{y} \right) \right\rangle $. This correlator depends on space only through its dependence on $\left\vert \vec{x} - \vec{y} \right\vert$ as a consequence of statistical isotropy and homogeneity. 

\begin{figure}[tbp]
	\centering 
	\includegraphics[width=0.4 \textwidth]{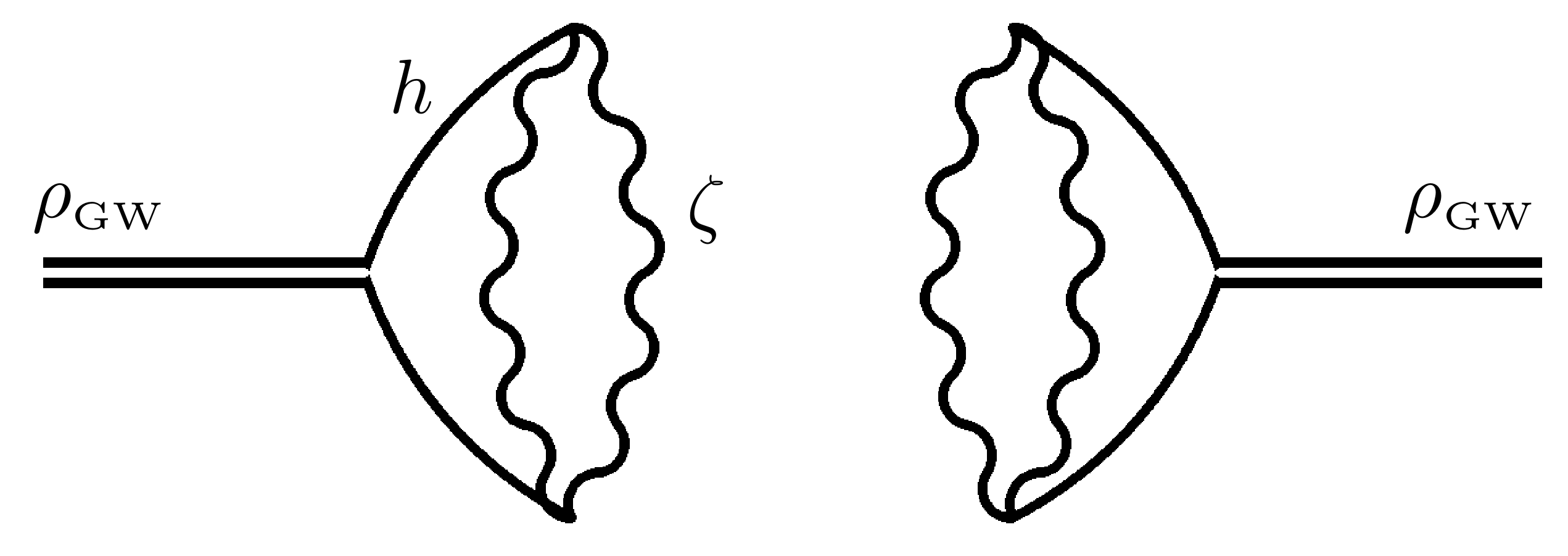}
	\caption{ Feynman diagram for the disconnected term in the energy density two-point function.
	The double lines identify the energy density field, the solid lines identify the gravitational waves and the wiggly lines identify the curvature field.
	}
	\label{fig:disconnected}
\end{figure}

In computing $\left\langle \rho_\GW^2 \right\rangle$ we need to evaluate the correlator $\left\langle \zeta^8 \right\rangle$. The resulting contractions are given in Eq. (\ref{wick}). The first line of that equation represents the case in which all the $\zeta$s emerging from the same $\rho$ are contracted among each other. This gives rise to the disconnected diagram shown in Figure \ref{fig:disconnected}, which is evaluated to  
\begin{equation} 
\left\langle \rho_\GW \left( \vec{x} \right)  \rho_\GW \left( \vec{y} \right) \right\rangle \Big\vert_{\rm disconnected} = \left\langle \rho_\GW \right\rangle^2, 
\end{equation} 
and is homogeneous. 

The other lines of Eq. (\ref{wick}) are represented by the different topologies of connected diagrams shown in  Figure \ref{fig:diagrams}.   As we show in Appendix \ref{app:connected} these contributions are completely negligible at the distances $\vert \vec{x} - \vec{y} \vert$ of our interest.  Our goal is to compute the large scale anisotropies in the GW energies arriving on Earth. The angular anisotropies (unless we go to extremely large multipoles $\ell$) are obtained by comparing the energy density from points which are separated by  non-negligible fractions of the present horizon. These distances are much greater than the wavelengths responsible for the GWs formation, $k_* \, \vert \vec{x} - \vec{y} \vert \gg 1$. For Gaussian scalar perturbations, the enhanced scalar modes (the first term in Eq. (\ref{calP})) do not lead to statistical correlations on these cosmological scales. It is also  easy to check that, as imposed by the Equivalence Principle, the anisotropies decay at large distances as $(q/k_*)^2$.

Furthermore,  when we measure the GW energy density at some given angular scale we effectively coarse grain the GW energy density with a resolution related to that scale. This results in averaging an extremely large number of patches of size $k_*^{-1}$, and the resulting energy density becomes extremely homogeneous due to the central limit theorem.

We conclude that the effects that we have discussed so far lead to a homogeneous and isotropic distribution of $\rho_\GW$, up to the completely negligible contributions from the terms evaluated in Appendix  \ref{app:connected}. There are however two additional effects of the long-scale modes that can lead to sizeable anisotropy. The first effect is a propagation effect \cite{Contaldi:2016koz,Bartolo:2019oiq}. Even if produced isotropically, GWs coming from different regions travel through disconnected and effectively different realizations of the large scales density perturbations. This makes the arriving GWs anisotropic. The second effect, that we study in this work, is that the scalar perturbations are not perfectly Gaussian. Most shapes of non-Gaussianity, starting from the most common local-type, give rise to correlations between short and long scales. Due to this, long wavelength modes can modulate the power of scales $k_*^{-1}$, thus giving rise to long-scale correlations in the initial GWs distribution. We study this effect in the next section.

\section{Primordial scalar non--Gaussianity and the angular anisotropies of the SGWB.} 
\label{sec:fNL} 

\begin{figure}[tbp]
	\centering 
	\includegraphics[width=0.4 \textwidth]{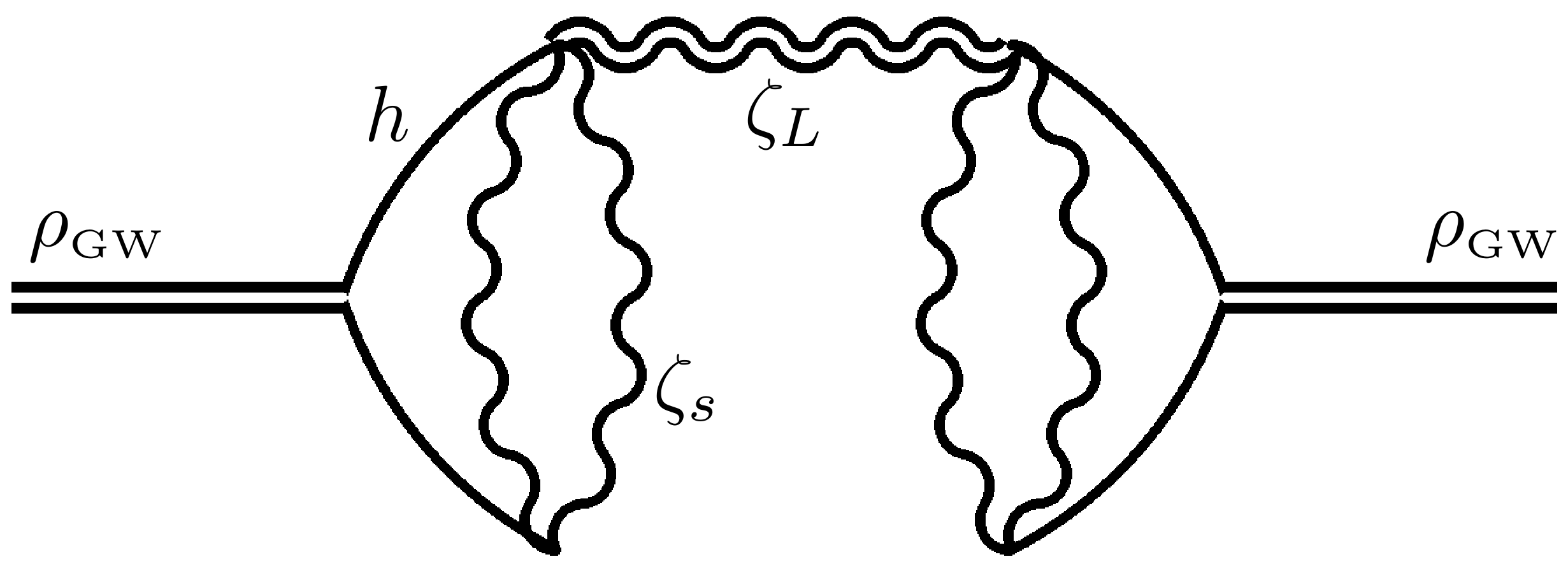}
	\caption{Feynman diagram for the energy density two-point function connected by a $f_\NL$ bridge. The double wiggly line indicates a $\zeta_L$ long mode.
	}
	\label{fig:connected}
\end{figure}

The non-Gaussianity of the primordial scalar perturbations is parametrized by 
\begin{equation}
\zeta( \vec{k}) = \zeta_g( \vec{k}) + \frac{3}{5} \, f_\NL \, \int \frac{d^3 p}{\left( 2 \pi \right)^3} \, 
\zeta_g \left( \vec{p} \right) \,  \zeta_g ( \vec{k} - \vec{p}),
\label{fNL-def}
\end{equation}  
namely it is assumed (as verified experimentally) that the perturbations are very close to be gaussian, so that a mode can be expanded as a large Gaussian contribution $ \zeta_g $ plus the square of a Gaussian term. The specific shape  in (\ref{fNL-def}) is known as local shape, as it corresponds to the local expansion $\zeta = \zeta_g + \frac{3}{5} f_\NL \, \zeta_g^2$ in real space. This is the most studied shape of non-Gaussianity, and it leads to significant correlation between large and small scales. Other shapes could also be considered, corresponding to a momentum-dependent non-linear parameter in the convolution (\ref{fNL-def}). For simplicity, in this work we consider only the local shape (\ref{fNL-def}). The Planck collaboration \cite{Akrami:2019izv} constrained the local non-linear parameter to 
\begin{equation}
- 11.1 \leq f_\NL \leq 9.3 \;\;\;,\;\;\; {\rm at} \; 95\% \; {\rm C.L.} 
\label{fNL-bound}
\end{equation} 
At the diagrammatic level, computing the two-point function using (\ref{fNL-def}) results in adding trilinear vertices $\zeta^3$, each proportional to $f_\NL$. In particular, vertices involving two short-scale and one long-scale mode connect the disconnected diagram of Figure \ref{fig:disconnected} into the connected diagram of Figure \ref{fig:connected}. For brevity, we will denote this connection as an ``$f_\NL$ bridge''.
In the peak-background split picture, one can expand the Gaussian comoving curvature perturbation field $\zeta_g$ as the sum of a short $\zeta_s$ and a long $\zeta_L$  components. In such a case, the four-point function in Eq. \eqref{4zeta}, not yet averaged over the long modes, results in
\begin{equation}
{\rm Non \  Gaussian \ \zeta} \rightarrow \la \zeta^4 \ra = \lp 1 + \frac{24}{5}f_\NL \zeta_L \rp \lp  \la \zeta_s^2 \ra \la \zeta_s^2 \ra + \la \zeta_s^2 \ra \la \zeta_s^2 \ra +\la \zeta_s^2 \ra \la \zeta_s^2 \ra \rp.
\end{equation}
In practice, it is convenient to write down the energy density before correlating over the long modes as 
\begin{equation}
	\rho_\GW (\eta, \vec{x}) = {\bar \rho_\GW} (\eta) \left[ 1 + \frac{24}{5} \, f_\NL 
	\int \frac{d^3 q}{\left( 2 \pi \right)^3} \, {\rm e}^{i \vec{q} \cdot \vec{x}}  \, \zeta_L \left( \vec{q} \right)  \right] , 
\end{equation} 
where the term ${\bar \rho_\GW}$ defines the energy density field at zeroth order in the non-linear parameter, while the second term in the square brackets accounts for the presence of such a non-Gaussianity.
From the energy density one can immediately compute the GWs abundance as
\begin{equation}
	\Omega_\GW ( \eta, \vec{x}, k) = {\bar \Omega_\GW} \left( \eta ,\, k \right)  
	\left[ 1 + \frac{24}{5} \, f_\NL \int \frac{d^3 q}{\left( 2 \pi \right)^3} \, {\rm e}^{i \vec{q} \cdot \vec{x}} \, \zeta_L \left( \vec{q} \right)  \right],
\end{equation}
where the term ${\bar \Omega_\GW} \left( \eta ,\, k \right)$ identifies the contribution with the absence of the long mode, see Eq. \eqref{monopole}.

\begin{figure}[t!]
	\centering 
	\includegraphics[width=0.49 \textwidth]{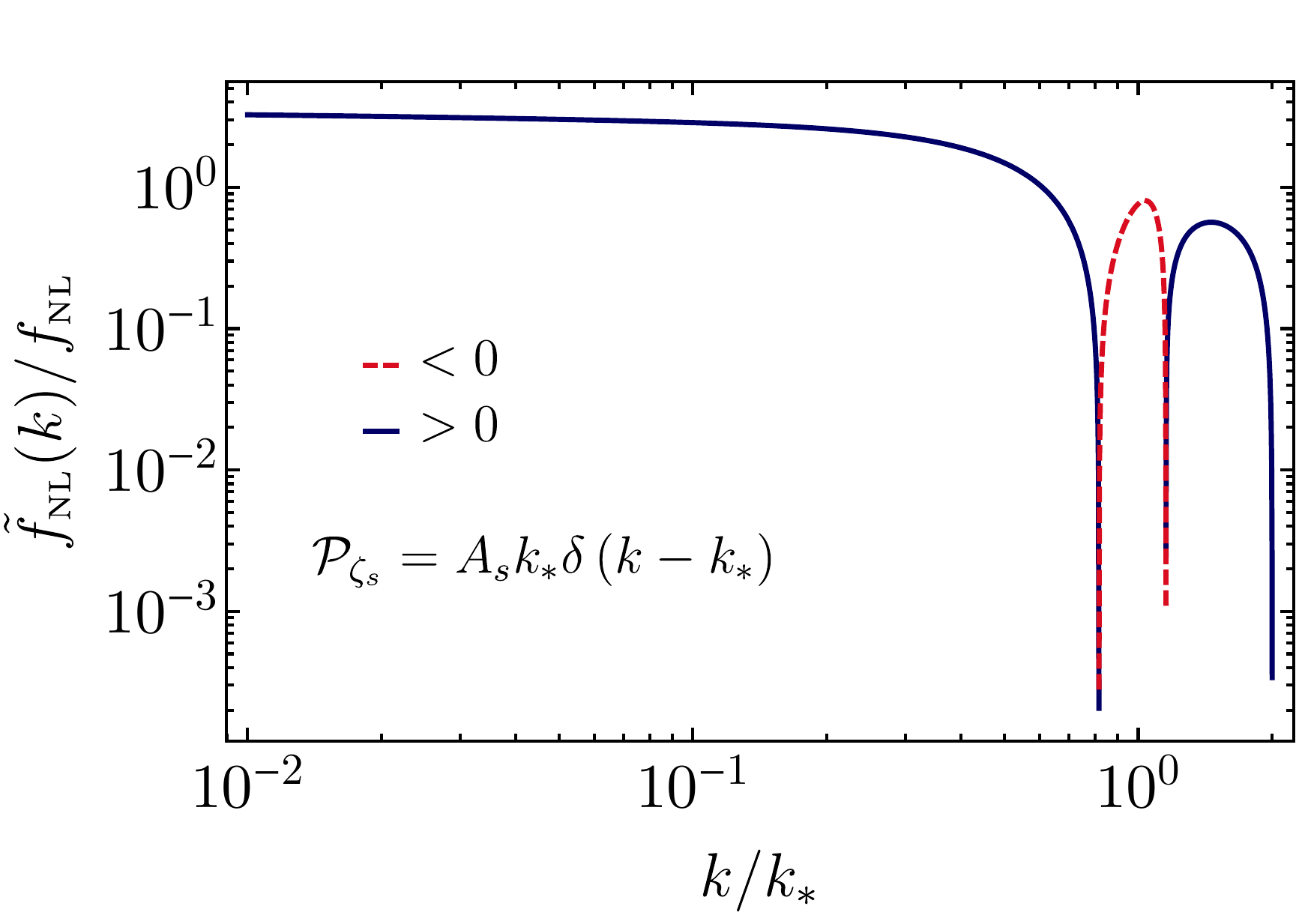}
	\includegraphics[width=0.49 \textwidth]{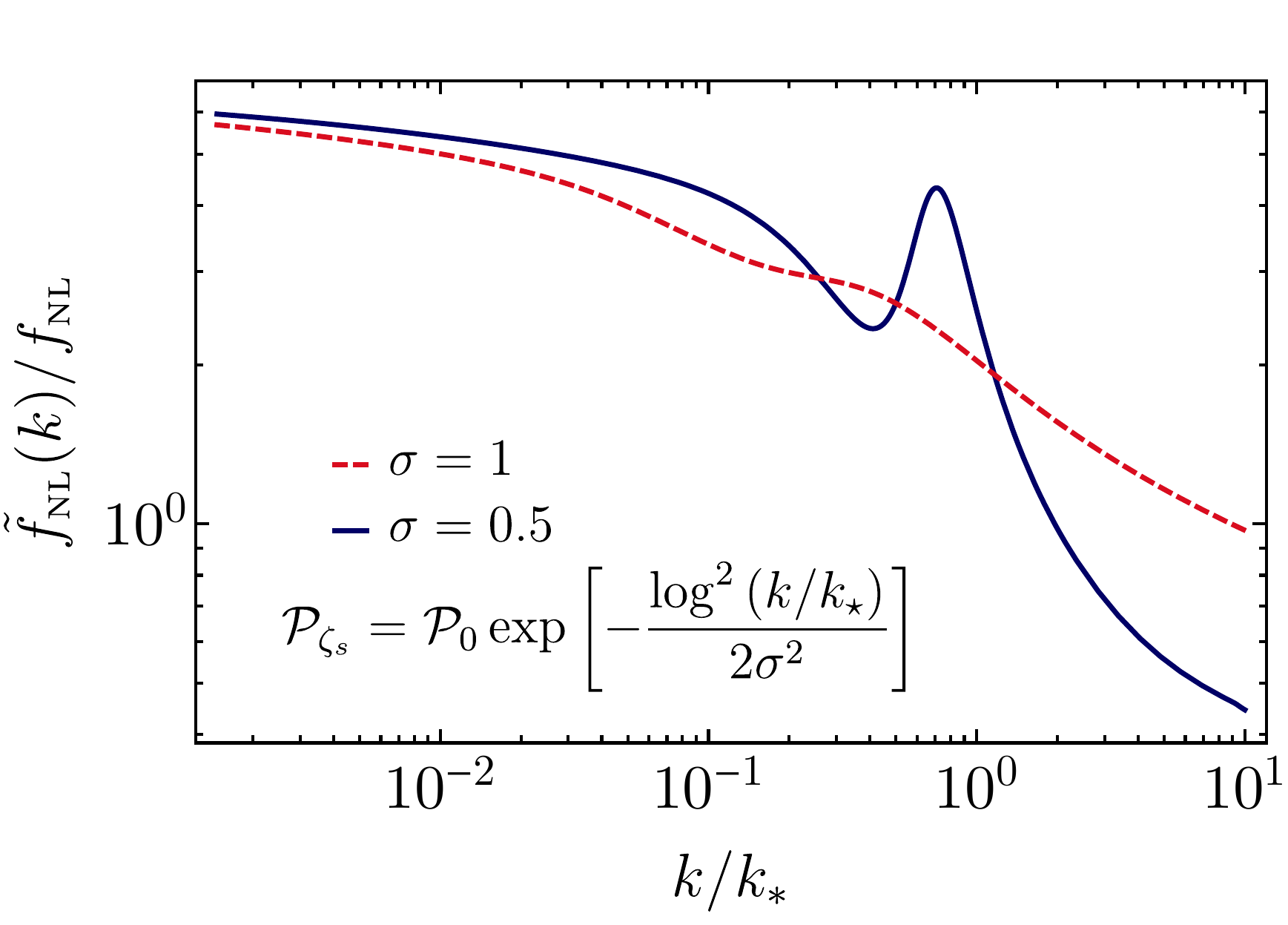}
	\caption{ ${\tilde f}_\NL/ f_\NL$ as a function of the ratio $k/k_*$ for a Dirac delta and gaussian power spectrum, respectively.	}
	\label{fig:fnltilde}
\end{figure}

Following the notation in \cite{Bartolo:2019oiq}, one can estimate the amount of anisotropy in the GW abundance by introducing the contrast
\begin{equation}
	\delta_\GW ( \eta, \vec{x}, \vec k)  = \frac{\Omega_\GW( \eta ,\, \vec{x} ,\, \vec  k) -  {\bar \Omega_\GW} \left( \eta ,\, k \right)  }{ {\bar \Omega_\GW} \left( \eta ,\, k \right)  } \equiv
\Gamma_I ( \eta, \vec{x}, \vec k) \lp 4-\frac{\partial \ln {\bar \Omega}_{\GW} (\eta ,\, k ) }{\partial \ln k} \rp	,
 \end{equation} 
in terms of the quantity 
 \begin{eqnarray} 
	\Gamma_I ( \eta, \vec{x}, \vec{k}) 
	=  \frac{3}{5} {\tilde f}_\NL \left( k \right)  \, \int \frac{d^3 q}{\left( 2 \pi \right)^3} \, {\rm e}^{i \vec{q} \cdot \vec{x}} 
	 \, \zeta_L \left( \vec{q} \right)  
,
	\qquad   {\tilde f}_\NL \left( k \right) \equiv  \frac{8 \, f_\NL}{4-\frac{\partial \ln {\bar \Omega}_{\GW} (\eta ,\, k ) }{\partial \ln k}}.
	\label{GammaI-time}
\end{eqnarray} 
This term carries all the information about the amount of anisotropy due to the initial conditions (suffix $I$). We choose to define the variable $\Gamma$ by following the notation used in \cite{Bartolo:2019oiq} where the subsequent propagation of the GWs in a perturbed  FLRW universe was originally studied by solving the free Boltzmann equation (for a discussion on the graviton collisional corrections see \cite{collisional} and Refs. therein).
Fig.~\ref{fig:fnltilde} shows the behaviour of the rescaled non-linear parameter as a function of the GW momentum for the choice of a Dirac delta and gaussian power spectrum.

Setting our location at the origin and defining $\vec{k} = k \, {\hat n}$,  then the position of the source term is at $\vec x = \hat n (\eta_{\rm in}-\eta)$, where $\eta_{\rm in}$ indicates the emission time which we associate to the moment when the modes $k_*$ re-enter the horizon and give rise to the signal we are considering in this work. One can expand this quantity using the spherical harmonics, to get
\footnote{We are using the spherical harmonics normalised as $\int d \hat n\, Y_{\ell m} Y_{\ell' m'}^* = \delta_{\ell \ell'} \delta_{mm'}$.}
\begin{eqnarray} 
	\Gamma_{\ell m, I} \left( k \right) 
	& = & 4 \pi \left( - i \right)^\ell \, \frac{3}{5} {\tilde f}_\NL \left( k \right) \, \int \frac{d^3 q}{\left( 2 \pi \right)^3} 
	\zeta_L \left( \vec{q} \right) \,  Y_{\ell m}^* \left( {\hat q} \right) \, j_\ell \left( q \left( \eta_0 - \eta_{\rm in} \right) \right) .
\end{eqnarray} 
To keep into account all the possible sources of anisotropy in the GW background, one shall add to this term the contribution from the propagation across the universe, 
\begin{equation}\label{gamma-propagation}
	\Gamma_{S}(\eta_0, \vec q) =  {\cal T}^S \left(  q ,\, \eta_0 ,\, \eta_{\rm in} \right) \zeta_L(\vec q),
\end{equation} 
where 
\begin{equation}
	{\cal T}^S \left(  q ,\, \eta_0 ,\, \eta_{\rm in} \right) = \int_{\eta_{\rm in}}^{\eta_0} d \eta' e^{-i \hat{k} \cdot {\hat q} q (\eta_0 - \eta')} \llp T_\Phi \left( \eta' ,\, q \right) \, \delta \left( \eta' - \eta_{\rm in} \right)  +
	\frac{\partial \left[ T_\Psi \left( \eta' ,\, q \right) +  T_\Phi \left( \eta' ,\, q \right) \right] }{\partial \eta'}  \rrp ,
\end{equation} 
and where 
\begin{equation} 
	\Phi ( \eta ,\, \vec{k}) \equiv T_\Phi( \eta ,\, k) \zeta( \vec{k}), \qquad 
	\Psi ( \eta ,\, \vec{k}) \equiv T_\Psi( \eta ,\, k) \zeta( \vec{k}).
\end{equation} 
The large scale modes of our interest entered the horizon during matter domination, and so the transfer functions become $ T_\Phi \left( \eta_{\rm in} ,\, q \right) = T_\Psi \left( \eta_{\rm in} ,\, q \right) =3/5$. 
Eq.~\eqref{gamma-propagation} represents the contribution of the scalar sources ($S$) when the signal travels across the universe towards us, and we see that it is composed by two pieces equivalent to the Sachs-Wolfe and integrated Sachs-Wolfe effect, respectively.

Therefore adding these two contributions one gets the full source of anisotropy as
\begin{equation}
	\Gamma_{\ell m,I+S} \left( k \right) = 4 \pi \left( - i \right)^\ell \,  
	\int \frac{d^3 q}{\left( 2 \pi \right)^3} \, 
	\zeta_L \left( \vec{q} \right) \, Y_{\ell m}^* \left( {\hat q} \right) \, {\cal T}_\ell^{I+S} \left(  k ,\, q ,\, \eta_0 ,\, \eta_{\rm in} \right) 
\end{equation}
where we defined the quantity
\begin{eqnarray}
	{\cal T}_\ell^{I+S} \left( k ,\, q ,\, \eta_0 ,\, \eta_{\rm in} \right) &\equiv & 
	\frac{3}{5} \left[ 1 + {\tilde f}_\NL \left( k \right)  \right]
  j_\ell \left( q \left( \eta_0 - \eta_{\rm in} \right) \right) \nonumber\\ 
  && + \int_{\eta_{\rm in}}^{\eta_0} d \eta \, 
	\frac{\partial \left[ T_\Psi \left( \eta ,\, q \right) +  T_\Phi \left( \eta ,\, q \right) \right] }{\partial \eta} \, 
	j_\ell \left( q \left( \eta_0 - \eta \right) \right). 
	\label{Gamma-IS}
\end{eqnarray}
Before going into the details of the computation of the correlators, one can have a deeper look of the ISW contribution to estimate its value with respect to the other. Introducing the variable $\eta' = \eta/\eta_0$ 
 and 
parametrising the scalar transfer functions as
\begin{equation}
T_\Phi \left( \eta ,\, q \right) = T_\Psi \left( \eta ,\, q \right) = \frac{3}{5} g(\eta),
\end{equation}
one gets
\begin{equation}
{\cal T}_\ell^S \left(k,  q ,\, \eta_0 ,\, \eta_{\rm in} \right) = 
 \frac{3}{5} \llp \left[ 1 + {\tilde f}_\NL \left( k \right)  \right] j_\ell \left( q \eta_0 \right)  + 2\int_{0}^{1} d \eta' \, 
\frac{\partial g(\eta') }{\partial \eta'} \, 
j_\ell \left( q \eta_0 (1-\eta') \right)  \rrp,
\end{equation}
where we neglected the term $q \eta_{\rm in}$ in the Bessel function of the first term.
Starting from the expression of $g(\eta)$, see for example Ref.~\cite{g1,g2}, one can use the analytical fit given by \cite{long}
\begin{equation}
\frac{\partial g (\eta')}{\partial \eta'} = -1.25 \eta'^5
\end{equation}
to perform the integral numerically, finding that the ISW effect is subdominant.
Therefore one can approximate the total contribution of the long mode, at leading order in the non-linear parameter, through the quantity
\begin{eqnarray}
\Gamma_{\ell m,I+S} \left( k \right) \simeq  4 \pi \left( - i \right)^\ell \,  
\int \frac{d^3 q}{\left( 2 \pi \right)^3} \, 
\zeta_L \left( \vec{q} \right) \, Y_{\ell m}^* \left( {\hat q} \right) \,  \frac{3}{5} 
\left[ 1 +  {\tilde f}_\NL \left( k \right) \right]  \, j_\ell \left( q \left( \eta_0 - \eta_{\rm in} \right) \right).
\label{Gml-IS-1}
\end{eqnarray}
In the following subsections we will compute the two-point and three-point functions of the rescaled energy density as a function of the long modes power spectra and the local non-linear parameter.

\subsection{Two-point function}
We start with the computation of the  two-point function 
\begin{align}
& \left \langle  \Gamma_{\ell_1 m_1, I+S} \left(k \right)  \Gamma^*_{\ell_2 m_2, I+S} \left(k \right) \right \rangle  = 
\left( 4 \pi \right)^2 \left( - i \right)^{\ell_1-\ell_2} \int \frac{d^3 q_1}{\left( 2 \pi \right)^3}  \frac{d^3 q_2}{\left( 2 \pi \right)^3} 
Y_{\ell_1 m_1}^* \left( {\hat q}_1 \right) Y_{\ell_2 m_2} \left( {\hat q}_2 \right) \nonumber\\ 
&
\times 
\left( \frac{3}{5} \right)^2 \left[ 1 +  {\tilde f}_\NL \left( k \right) \right]^2  
j_{\ell_1} \left( q_1 \left( \eta_0 - \eta_{\rm in} \right) \right) 
j_{\ell_2} \left( q_2 \left( \eta_0 - \eta_{\rm in} \right) \right) 
\left\langle \zeta_L \left( \vec{q}_1 \right) \zeta_L^* \left( \vec{q}_2 \right) \right\rangle .
\end{align}
Using the orthonormality of the spherical harmonics and for the choice of a scale invariant power spectra of the long modes ${\cal P}_{\zeta_L} (q) = {\cal P}_{\zeta_L}$, the previous expression becomes 
\begin{align}
\left \langle  \Gamma_{\ell_1 m_1, I+S} \left(k \right)  \Gamma^*_{\ell_2 m_2, I+S} \left(k \right) \right \rangle = \delta_{\ell_1 \ell_2} \delta_{m_1 m_2} 4 \pi \lp \frac{3}{5}  \rp^2 \left[ 1 + {\tilde f}_\NL(k) \right]^2
\frac{1}{2 \ell_1 \left( \ell_1 + 1 \right)}  {\cal P}_{\zeta_L}.
\end{align}
Following the notation of \cite{Bartolo:2019oiq}, one can define the two-point function as
\begin{equation}
\left \langle  \Gamma_{\ell_1 m_1, I+S} \left(k \right)  \Gamma^*_{\ell_2 m_2, I+S} \left(k \right) \right \rangle = \delta_{\ell_1 \ell_2} \delta_{m_1 m_2} \, C_{\ell,I+S} \left( k \right) 
\end{equation} 
such that one finally gets
\begin{eqnarray} 
\label{ClGamma}
\sqrt{\frac{\ell \left( \ell+1 \right)}{2 \pi} \, C_{\ell,I+S} \left( k \right)}  &\simeq& \frac{3}{5} 
\left\vert 1 + {\tilde f}_{\NL} \left( k \right) \right\vert \, {\cal P}_{\zeta_L}^{1/2} \simeq 2.8 \cdot 10^{-4} \, \left\vert \frac{ 1 + {\tilde f}_{\NL} \left( k \right) }{10} \right\vert \, \left( \frac{{\cal P}_{\zeta_L}}{2.2 \cdot 10^{-9}} \right)^{1/2} \nonumber \\
\end{eqnarray} 
which has been evaluated for value of the non-linear parameter close to its upper bound (\ref{fNL-bound}) and using the CMB value for the power spectrum of the long modes.

\subsection{Three-point function}
For the computation of the three-point function we need to go to the next-to-leading order in the non-linear parameter $f_\NL$, such that the expression of the initial condition term $\Gamma_I$ in the $\ell,m$ space  becomes
\begin{align}
	\Gamma_{\ell m,I} \left( k \right) &\simeq   4 \pi \left( - i \right)^\ell \,  
	\int \frac{d^3 q}{\left( 2 \pi \right)^3} \, 
	 \, Y_{\ell m}^* \left( {\hat q} \right) \,   \, j_\ell \left( q \left( \eta_0 - \eta_{\rm in} \right) \right) 
	 \frac{3}{5} 
	{\tilde f}_\NL \left( k \right)
	\left [ \zeta_L \left( \vec{q} \right) +  \frac{9}{5} \, f_\NL  \int \frac{d^3 p}{\left( 2 \pi \right)^3}
	\,  \zeta_L \left( \vec{p} \right)     \zeta_L \left(\vec{q}-\vec p \right)  \right ].
	\nonumber\\ 
\end{align}
At this order in the long perturbations $\zeta_L $, also the propagation term gets a contribution proportional to the non-linear parameter as $3/5 f_\NL \zeta^2_L$, so that the total  term becomes
\begin{eqnarray}
\Gamma_{\ell m,I+S} \left( k \right) &\simeq&   4 \pi \left( - i \right)^\ell \,  
\int \frac{d^3 q}{\left( 2 \pi \right)^3} \, 
\, Y_{\ell m}^* \left( {\hat q} \right) 
\, j_\ell \left( q \left( \eta_0 - \eta_{\rm in} \right) \right) \, \Bigg\{  \frac{3}{5} \left[ 1 + {\tilde f}_\NL \left( k \right) \right] \zeta_L \left( \vec{q} \right) \nonumber\\ 
&+&    \frac{9}{25} \, f_\NL \left[ 1 + 3  {\tilde f}_\NL \left( k \right)  \right]  \, \int \frac{d^3 p}{\left( 2 \pi \right)^3}
\,  \zeta_L \left( \vec{p} \right)     \zeta_L \left(\vec{q}-\vec p \right) \Bigg\} 
\end{eqnarray}
where we stress once again that all the long modes $\zeta_L$ in this expression are Gaussian fields.
We can now start the evaluation of the three-point function
\begin{align}
&
\left\langle \prod_{i=1}^3 \Gamma_{\ell_i m_i,I+S} \left( k \right) \right\rangle = 
\left( 4 \pi \right)^3 \left( - i \right)^{\ell_1+\ell_2+\ell_3} 
\frac{81}{625} \, f_\NL \, \left[ 1 +  {\tilde f}_\NL \left( k \right) \right]^2 \, \left[ 1 + 3 {\tilde f}_\NL \left( k \right) \right] 
\int \frac{d^3 q_1}{\left( 2 \pi \right)^3} \int \frac{d^3 q_2}{\left( 2 \pi \right)^3} 
 \nonumber\\ 
&
\times \int \frac{d^3 q_3}{\left( 2 \pi \right)^3} 
\int \frac{d^3 p}{\left( 2 \pi \right)^3}
\left[ \prod_{i=1}^3 Y_{\ell_i m_i}^* \left( {\hat q}_i \right) \,  j_{\ell_i} \left( q_i \left( \eta_0 - \eta_{\rm in} \right) \right) \right] 
\left\langle \zeta_L \left( \vec{q}_1 \right)  \zeta_L \left( \vec{q}_2 \right)  \zeta_L \left( \vec{p} \right)  \zeta_L \left( \vec{q}_3 - \vec{p} \right) \right\rangle + 2 \, {\rm perm.}
\end{align}
After having performed the contractions of the long modes with the Wick theorem, one can introduce the bispectrum of the modes in momentum space ${\cal B}_\Gamma$, such that the previous expression becomes
\begin{align}
\left\langle \prod_{i=1}^3 \Gamma_{\ell_i m_i,I+S} \left( k \right) \right\rangle &= 
\left( 4 \pi \right)^3 \left( - i \right)^{\ell_1+\ell_2+\ell_3} 
\int \frac{d^3 q_1}{\left( 2 \pi \right)^3} \int \frac{d^3 q_2}{\left( 2 \pi \right)^3} \int \frac{d^3 q_3}{\left( 2 \pi \right)^3}  
{\cal B}_\Gamma \left( k ,\, q_1 ,\, q_2 ,\, q_3 \right)  \nonumber\\ 
& \times
\left[ \prod_{i=1}^3 Y_{\ell_i m_i}^* \left( {\hat q}_i \right) \,  j_{\ell_i} \left( q_i \left( \eta_0 - \eta_{\rm in} \right) \right) \right] 
\left( 2 \pi \right)^3 \delta^{(3)} \left( \vec{q}_1 + \vec{q}_2 + \vec{q}_3  \right) 
\end{align} 
with 
\begin{equation}
{\cal B}_\Gamma \left( k ,\, q_1 ,\, q_2 ,\, q_3 \right) = \frac{162}{625} \, f_\NL \left[ 1 +  {\tilde f}_\NL \left( k \right) \right]^2 \, \left[ 1 + 3 {\tilde f}_\NL \left( k \right) \right] \left[ 
\frac{2 \pi^2}{q_1^3} {\cal P}_{\zeta_L} \left( q_1 \right) 
\frac{2 \pi^2}{q_2^3} {\cal P}_{\zeta_L} \left( q_2 \right) 
+ 2 \, {\rm perm.}  
\right].
\end{equation} 
Using the representation of the Dirac $\delta$-function in terms of the spherical harmonics, and using their orthonormality, one gets after some algebra
\begin{align} 
\left\langle \prod_{i=1}^3 \Gamma_{\ell_i m_i,I+S} \left( k \right)  \right\rangle =  {\cal G}_{\ell_1 \ell_2 \ell_3}^{m_1 m_2 m_3} \, \int_0^\infty d r \, r^2 \,  \prod_{i=1}^3 \left[ \frac{2}{\pi} \int d q_i \, q_i^2 j_{\ell_i} \left( q_i \left( \eta_0 - \eta_{\rm in} \right) \right) \, j_{\ell_i} \left( q_i \, r \right) \right] \, {\cal B}_\Gamma \left( k,\, q_1 ,\, q_2 ,\, q_3 \right) \nonumber\\ 
\end{align} 
where one could recognize the Gaunt integral 
\begin{equation}
{\cal G}_{\ell_1 \ell_2 \ell_3}^{m_1 m_2 m_3} =  \int d \Omega_y \, Y_{\ell_1 m_1}^* \left( \Omega_y \right) \, Y_{\ell_2 m_2}^* \left( \Omega_y \right) \, Y_{\ell_3 m_3}^* \left( \Omega_y \right).
\end{equation}
In the limit of one sufficiently large $\ell$, it is then possible to evaluate one of the $q$ integral, for each of the three permutations in the bispectrum, by using the approximation 
\begin{equation} 
\frac{2}{\pi} \, \int d q \, q^2 j_{\ell} \left( q \, \eta_0  \right) \, j_{\ell} \left( q \, r \right) \bigg | _{\ell \gg 1}= \frac{\delta \left( \eta_0 - r \right)}{\eta_0^2} 
\end{equation}
and then use the resulting Dirac delta to integrate over $r$. The result of this computation is therefore
\begin{align} 
&
\left\langle \prod_{i=1}^3 \Gamma_{\ell_i m_i,I+S} \left( k \right)  \right\rangle =  {\cal G}_{\ell_1 \ell_2 \ell_3}^{m_1 m_2 m_3} \, 
\frac{162}{625} \, f_\NL \, \left[ 1 +  {\tilde f}_\NL \left( k \right) \right]^2 \, \left[ 1 + 3 {\tilde f}_\NL \left( k \right) \right] \nonumber\\ 
& \quad\quad\quad\quad \times 
\lp 4\pi \, \int \frac{d q_1}{q_1} j_{\ell_1}^2 \left( q_1 \, \eta_0  \right) {\cal P}_{\zeta_L} \left( q_1 \right) \rp \, 
\lp 4 \pi \int \frac{d q_2}{q_2} j_{\ell_2}^2 \left( q_2 \, \eta_0  \right) {\cal P}_{\zeta_L} \left( q_2 \right) \rp
+ 2 \, {\rm perm.}
\end{align} 
Finally, one can factorize the tensorial structures following from statistical isotropy to define the three-point function as \cite{Bartolo:2004if}
\begin{eqnarray}
 \left\langle \Gamma_{\ell_1 m_1,I+S}(k)   \Gamma_{\ell_2 m_2,I+S}(k)  \Gamma_{\ell_3 m_3,I+S}(k)  \right\rangle =  {\cal G}_{\ell_1 \ell_2 \ell_3}^{m_1 m_2 m_3} \, {b}_{\ell_1 \ell_2 \ell_3,I+S} \left( k \right) ,
\end{eqnarray} 
where, in terms of the two-point functions found in Eq. \eqref{ClGamma}, the expression becomes
\begin{eqnarray}
{ b}_{\ell_1 \ell_2 \ell_3,I+S} \left( k \right) &\simeq& 
\frac{ 2 \, f_\NL \, \left[ 1 + 3 \, {\tilde f}_\NL \left( k \right) \right]}{  \left[ 1 +  {\tilde f}_\NL \left( k \right) \right]^2 } 
\left[ C_{\ell_1,I+S} \,  C_{\ell_2,I+S} +  C_{\ell_1,I+S} \,  C_{\ell_3,I+S} +  C_{\ell_2,I+S} \,  C_{\ell_3,I+S} \right].
\nonumber\\ 
\label{summary-Gamma}
\end{eqnarray} 
We dedicate the next sections to the discussion of these results.

\section{Short summary of the isocurvature constraints on non-Gaussianity}
\label{sec:iso}

The presence of such a non-Gaussianity in the comoving curvature perturbation has a small effect 
on the value of the threshold which is necessary to the overdensity to collapse into PBHs, see Ref. \cite{Kehagias:2019eil} for details, while it induces a significant large-scale variation of  the primordial black holes abundance through the modulation of the power on small scales induced by the long modes. 
If all or a part of the dark matter is composed by PBHs, then this non-Gaussianity is responsible for the production of isocurvature modes in the DM density fluid, which are strongly constrained by the CMB observations.

The present bounds provided by the Planck experiments on the relative abundance of the isocurvature modes are, at 95\% CL, \cite{Akrami:2018odb}
\begin{align}
\label{bounds-beta}
& 100 \beta_{\rm iso} < 0.095 \qquad \text{ for fully  correlated}, \nonumber \\
& 100 \beta_{\rm iso} < 0.107 \qquad\  \text{for fully  anti-correlated},
\end{align}
where by fully correlated (fully anti-correlated) we mean a positive (negative) $f_\NL$.

Following the results obtained in \cite{Young:2015kda}, one can express the PBH mass fraction in the presence of non-Gaussianity. It reads (see also \cite{NGnoi})
\begin{equation}
\bar\beta
\equiv \frac{\rho_\PBH(\eta_{\rm in})}{\rho_{\rm c}(\eta_{\rm in})}
=
\left \{
\begin{aligned}
\sqrt{\frac{2}{\pi\sigma_s^2}}\left[\int_{\zeta_{+}}^\infty d\zeta\, {\rm exp}\left(-\frac{\zeta^2}{2\sigma_s^2}\right)+\int^{\zeta_{-}}_{-\infty} d\zeta\, {\rm exp}\left(-\frac{\zeta^2}{2\sigma_s^2}\right)\right]
\quad  {\rm for} \quad f_\NL>0,
\\
\sqrt{\frac{2}{\pi\sigma_s^2}}\left[\int_{\zeta_{+}}^\infty d\zeta\, {\rm exp}\left(-\frac{\zeta^2}{2\sigma_s^2}\right)-\int_{\zeta_{-}}^{\infty} d\zeta\, {\rm exp}\left(-\frac{\zeta^2}{2\sigma_s^2}\right)\right]
\quad  {\rm for} \quad
f_\NL<0,
\end{aligned}
\right. 
\end{equation}
where \cite{Young:2015kda}
\begin{equation}
\zeta_{\pm}=\frac{-5\pm\sqrt{25+60\zeta_c f_\NL+36 f_\NL^2\sigma_s^2}}{6 f_\NL}
\end{equation}
and $\zeta_c$ is the threshold for collapse of PBH in the presence of non-Gaussianity recently calculated in Ref. \cite{zetafnl}, and $\sigma_s^2$ is the variance of the short modes.

The corresponding mass fraction perturbation with respect to the average value 
$\bar \beta$ at leading order in the long modes is
\begin{equation}
\delta_\beta \equiv \frac{\beta - \bar \beta}{\bar \beta}
= \lp \frac{25 + 30 \zeta_c f_\NL +36 f_\NL^2 \sigma_s^2 - 5 \sqrt{25 + 60 \zeta_c f_\NL +36 f_\NL^2 \sigma_s^2}}{3 f_\NL \sigma_s^2\sqrt{25 + 60 \zeta_c f_\NL +36 f_\NL^2 \sigma_s^2} }
\rp \zeta_L
\equiv  b\,\zeta_L.
\end{equation}
One can express the relative abundance of the isocurvature modes in terms of the bias $b$ induced by the long mode as
\begin{equation}
\beta_{\rm iso} \equiv \frac{\mathcal{P}_{\rm iso}}{\mathcal{P}_{\rm iso} + \mathcal{P}_{\zeta_L}} = \frac{b^2 f_\PBH^2}{b^2 f_\PBH^2+ 1},
\end{equation}
where we used the fact that local non-Gaussianity induces the bias $\mathcal{P}_{\rm iso} = b^2 f_\PBH^2 \mathcal{P}_{\zeta_L}$, 
where $ f_\PBH$ is the fraction of dark matter in PBH. Once written in terms of $b$ and $f_\PBH$, the bounds (\ref{bounds-beta}) become
\begin{equation}
- 0.0327 < b f_\PBH < 0.0308.
\end{equation}
One can finally relate the bias to the parameter of non-Gaussianity as done in \cite{Young:2015kda}, giving  the colored allowed region in Fig. \ref{fig:plot2pt12Msun}.  In making the plot, we
are assuming that the value of local $f_{\NL}$ has no scale dependence, as explained in the Introduction. From the plot it is clear that a large value of the non-linear parameter implies that only a small fraction of DM can be composed by PBHs. We remind to the reader the fact that the non-linear parameter has a lower bound due to the inadequacy of the perturbative approach in the computation of the PBH abundance \cite{Young:2013oia, Yoo:2019pma}, because of which we decided to cut the allowed region in the plot at $f_{\NL} \geq -1/3$.

\section{Results}
\label{sec:results} 

To have a more physical intuition of the amount of anisotropy in the GWs abundance, we express the above results in terms of the GW density contrast $\delta_\GW$ rather than of $\Gamma$. We thus define the two and three point functions as
\begin{eqnarray}
	&& \left\langle \delta_{\GW,\ell m}  \delta_{\GW,\ell' m'}^* \right\rangle = \delta_{\ell \ell'} \delta_{m m'} \, {\hat C}_\ell \left( k \right), \nonumber\\ 
	&& \left\langle  \delta_{\GW,\ell_1 m_1}  \delta_{\GW,\ell_2 m_2}  \delta_{\GW,\ell_3 m_3}  \right\rangle =  {\cal G}_{\ell_1 \ell_2 \ell_3}^{m_1 m_2 m_3} \, {\hat b}_{\ell_1 \ell_2 \ell_3} \left( k \right), 
\end{eqnarray} 
where we have again factorised the tensorial structures dictated by statistical isotropy,
such that the above results then become 
\begin{eqnarray}
	\sqrt{\frac{\ell \left( \ell+1 \right)}{2 \pi} \, {\hat C}_{\ell} \left( k \right)}  &\simeq& \frac{3}{5} 
	\left\vert 1 + {\tilde f}_{\NL} \left( k \right) \right\vert \, \left\vert  4-\frac{\partial \ln {\bar \Omega}_{\GW} (\eta ,\, k ) }{\partial \ln k} \right\vert \,  {\cal P}_{\zeta_L}^{1/2}, \nonumber\\ 
{\hat b}_{\ell_1 \ell_2 \ell_3} \left( k \right) &\simeq& 
\frac{ \, \tilde{f}_\NL \, \left[ 1 + 3 \, {\tilde f}_\NL \left( k \right) \right]}{ 4 \left[ 1 +  {\tilde f}_\NL \left( k \right) \right]^2 } 
\lp {\hat C}_{\ell_1} \,  {\hat C}_{\ell_2} +  {\hat C}_{\ell_1} \, {\hat  C}_{\ell_3} + {\hat  C}_{\ell_2} \, {\hat  C}_{\ell_3} \rp.
\end{eqnarray} 
We can now discuss the limits in which the anisotropies are dominated by the propagation term or by the initial condition term. 

In the case in which the propagation term dominates, one can formally consider the limit ${\tilde f}_\NL \to 0$, and thus find
\begin{eqnarray}
	{\rm Dominated \; by \; propagation} : \left\{ \begin{array}{l} 
		\sqrt{\frac{\ell \left( \ell+1 \right)}{2 \pi} \, \hat C_{\ell} \left( k \right)}  \simeq \frac{3}{5} \, \left\vert  4-\frac{\partial \ln {\bar \Omega}_{\GW} (\eta ,\, k ) }{\partial \ln k} \right\vert \, {\cal P}_{\zeta_L}^{1/2},
		\\ \\ 
		{\hat b}_{\ell_1 \ell_2 \ell_3} \left( k \right) \simeq 
		\frac{ 1}{4}\tilde f_\NL \, \left[ \hat C_{\ell_1} \, \hat  C_{\ell_2} + \hat  C_{\ell_1} \, \hat  C_{\ell_3} +  \hat C_{\ell_2} \, \hat  C_{\ell_3} \right],
	\end{array} \right.
\end{eqnarray} 
which agrees with the results of the previous paper \cite{Bartolo:2019oiq}.

In the case in which the initial condition term dominates, one can instead consider the limit ${\tilde f}_\NL \to \infty$. The correlators for $\delta_\GW$ then become 
\begin{eqnarray}
	{\rm Dominated \; by \; initial \; condition} : \left\{ \begin{array}{l} 
		\sqrt{\frac{\ell \left( \ell+1 \right)}{2 \pi} \, {\hat C}_{\ell} \left( k \right)}  \simeq \frac{24}{5} \, \left\vert f_\NL \right\vert   \,  {\cal P}_{\zeta_L}^{1/2}, \\ \\ 
		{\hat b}_{\ell_1 \ell_2 \ell_3} \left( k \right) \simeq 
		\frac{ 3 }{ 4 } 
		\left( {\hat C}_{\ell_1} \,  {\hat C}_{\ell_2} +  {\hat C}_{\ell_1} \, {\hat  C}_{\ell_3} + {\hat  C}_{\ell_2} \, {\hat  C}_{\ell_3} \right),
	\end{array} \right. 
\end{eqnarray} 
where we note that $f_\NL$ has disappeared from the last expression, since $\Gamma_I$ is maximally non-Gaussian (as opposite to $\Gamma_S$, that is Gaussian up to ${\mathcal O } \left( f_\NL \right)$ non-Gaussianity).

\begin{figure}[tbp]
	\centering 
	\includegraphics[width=0.485 \textwidth]{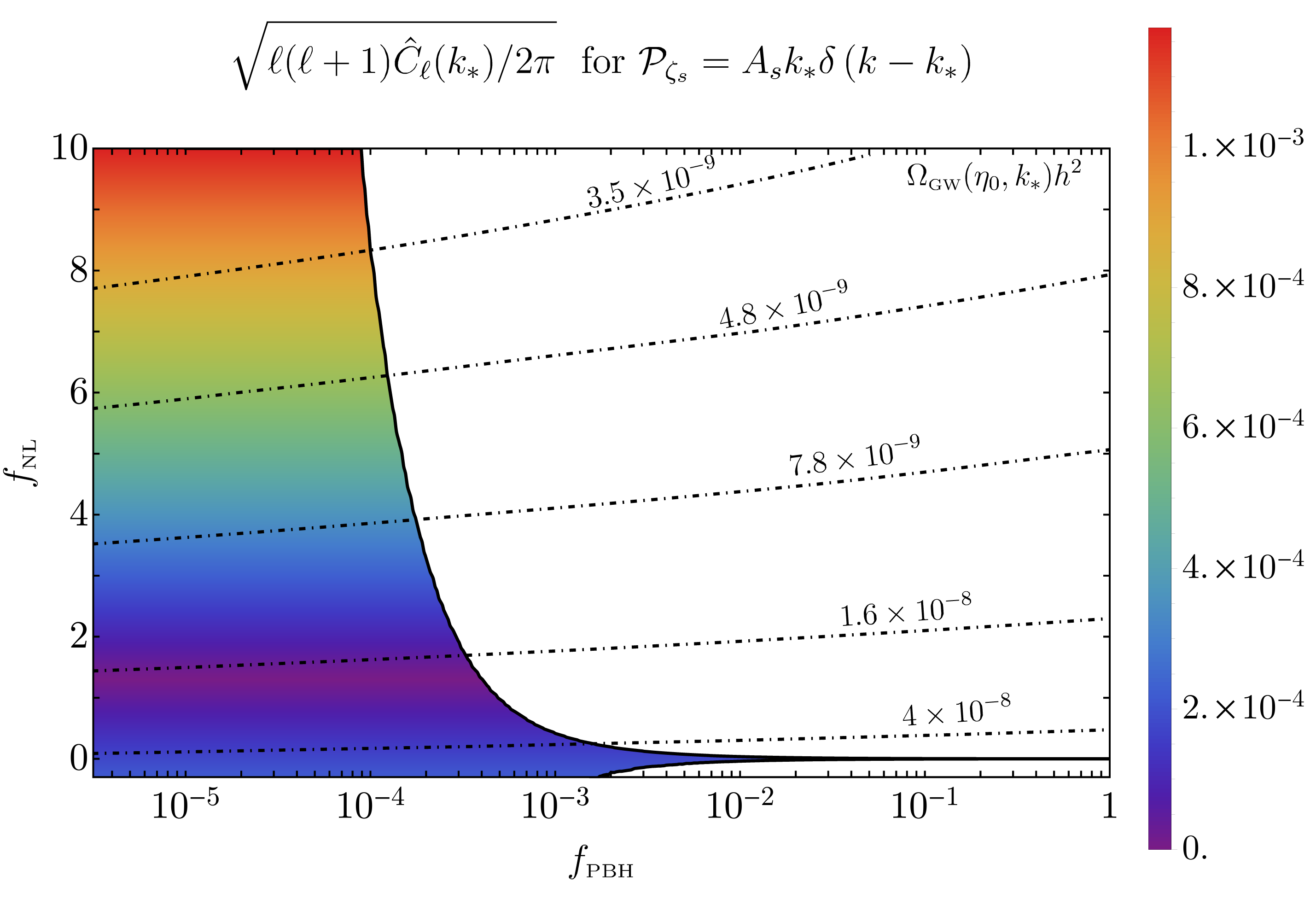}
	\includegraphics[width=0.495 \textwidth]{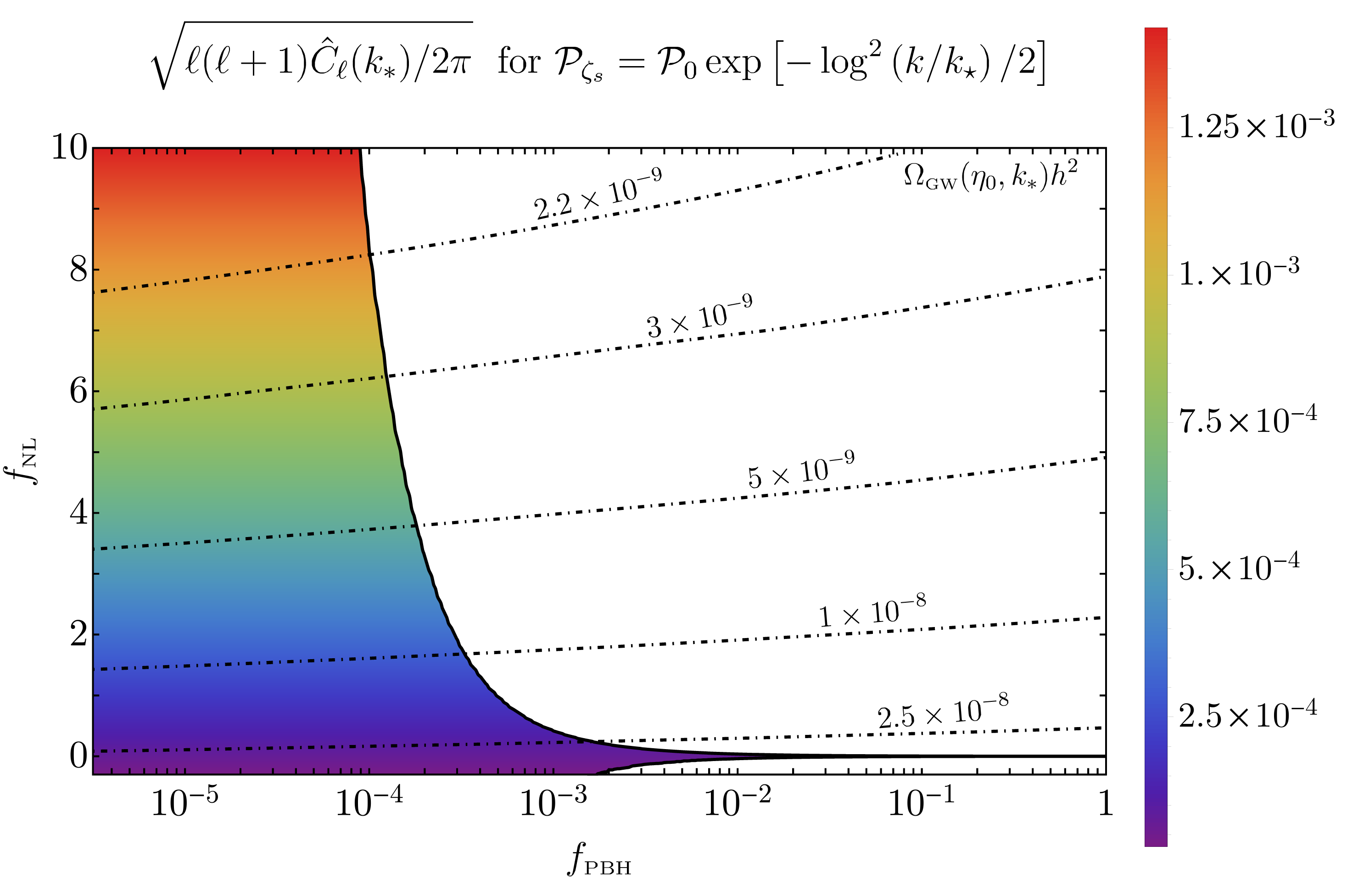}
	\caption{ Contour plot of $\sqrt{\ell (\ell + 1)\hat{C_\ell}(k_*)/2 \pi}$ in the region permitted by the constraints of Planck on $f_\PBH$ and $f_\NL$ for the choice of a Dirac delta and gaussian power spectrum of the short modes, respectively. The peak frequency has been chosen to correspond to $M_\PBH = 10^{-12} M_\odot$.
		The dot-dashed lines identify the corresponding GWs abundance.  Notice that the results shown here  only hold  for a local, scale-invariant primordial non-Gaussianity of scalar perturbations.}
	\label{fig:plot2pt12Msun}
\end{figure}

In Fig.~\ref{fig:plot2pt12Msun} we show the two-point function anisotropy $\hat{C_\ell}$ for the density contrast, for the choice of a Dirac delta and gaussian power spectrum of the curvature perturbation on small scales. The peak frequency of this signal has been chosen as the one corresponding to PBH masses given by $M_\PBH = 10^{-12} M_\odot$ for which PBHs can represent all the DM, also coinciding with the frequency of maximum sensitivity at LISA.
The dot-dashed lines identify the corresponding GWs abundance computed at present time and at the peak frequency.
Finally, the results for different masses of PBH do not change significantly.

\section{Conclusions}
\label{sec:conclusions} 
The measurement of a SGWB is one of the main goals of future experiments devoted to the detection of sources of GWs. One possible and well-motivated source of GWs from the early universe is associated to the birth of PBHs when enhanced scalar perturbations created during inflation re-enter the horizon and collapse into  BHs. This phenomenon is accompanied by the generation of GWs at second-order in perturbation theory. In particular, it turns out that for PBHs of masses around $10^{-12} M_\odot$, which can still play the role of dark matter in its totality, the frequency of the GWs is located 
in the mHz range where  the LISA mission happens to have the maximum sensitivity. In the positive case of a detection of the SGWB, the next step will be to identify the source and therefore any 
characterisation of the background will be extremely useful. In this sense, its anisotropies will bring important information. 

In this paper we have studied in detail  the strength of the GW anisotropies associated to the 
production of the PBHs. There are two contributions to the anisotropy, the first one is created at the generation epoch and the second one is due to the propagation effects from the time of production down to 
the detection time. In order to have the first source on large scales a non-vanishing  squeezed type of non-Gaussianity must be present in the curvature perturbation in order to create a cross-talk between the PBH short wavelengths and the large scales at which the anisotropies are tested. At the same time, the amount of primordial non-Gaussianity is constrained by the requirement of not generating a too large amount of isocurvature perturbations, in the case in which PBHs compose a sizeable fraction of the dark matter. 

 We have considered the simplest possibility, namely a primordial scale-invariant local non-Gaussianity for the curvature perturbations. Under such an assumption, our results  are summarised in Fig.~\ref{fig:plot2pt12Msun} out of which we conclude that the typical anisotropies are of the order of $\zeta_L\sim 10^{-4}$. Correspondingly, the reduced bispectrum is of the order of  $\zeta^2_L\sim 10^{-8}$. Our findings show also that, if the PBHs compose a large fraction of the dark matter, the SGWB must be highly isotropic and Gaussian, up to propagation effects.  A large amount of anisotropy and non-Gaussianity would imply, within our mechanism, a PBH population well below the measured dark matter abundance. 

Such conclusions hold only in the case of our working hypothesis, namely a local model of primordial non-Gaussianity with $f_{\rm NL}={\rm const}$ for the curvature perturbations. In this case, one is directly using the {\it Planck} constraints on $f_{\rm NL}$ (and the isocurvature limits discussed in Sec.~\ref{sec:iso})  down to the scales typical of PBH formation. However, if that is not the case, then our constraints shown in Fig.~\ref{fig:plot2pt12Msun} can be relaxed, with PBHs that might constitute all of the measured dark matter. For example, one possibility might be to extend our computation by considering a running (local) non-Gaussianity   \cite{Sefusatti:2009xu}, which is presently constrained by CMB temperature measurements \cite{ Oppizzi:2017nfy} and might possibly avoid the isocurvature bounds. We leave it for further studies.

The next step is of course understanding if such small anisotropies can be detected by the current and future experiments and, if so, at which angular resolution \cite{a}. In particular, for a SGWB of cosmological origin only anisotropies at  low multipoles, $\ell\lesssim 10$,  can be resolved. To  resolve  the angular features of the SGWB at  larger multipoles, 
a gravitational wave telescope characterised by  a $\sim$ AU effective baseline seems to represent  the best option \cite{a}.

\vskip.25cm
\section*{Acknowledgements} 
We thank C. Byrnes for useful discussions.
N.B., D.B. and S.M. acknowledge partial financial support by ASI Grant No. 2016-24-H.0.
V.DL., G.F. and A.R. are  supported by the Swiss National Science Foundation (SNSF), project {\sl The Non-Gaussian Universe and Cosmological Symmetries}, project number: 200020-178787.
The work of G.T. is partially supported by STFC grant ST/P00055X/1.

\vskip.25cm

\appendix

\section{Conventions and computational details on the SGWB energy density}
\label{app:conventions}

In this Appendix we list our conventions and some explicit expressions that are relevant for the GW energy density in eq. (\ref{rhox}). 
We introduce the GW field through the line element 
\begin{equation}
ds^2 = a^2(\eta) \left[ -d\eta^2 +\left( \delta_{ij} +h_{ij} \right) dx^i  dx^j \right],
\end{equation}
and we decompose it as 
\begin{equation}
h_{ij} \left( \eta ,\, \vec{x} \right) = \int \frac{d^3 k}{\left( 2 \pi \right)^3} \sum_{\lambda = R,L} h_\lambda ( \eta ,\, \vec{k} )\,   e_{ij,\lambda}  ({\hat k} ) \, {\rm e}^{i \vec{k} \cdot \vec{x} }, 
\label{h-sol}
\end{equation} 
where the circular polarization operators are transverse and traceless, and satisfy the normalization condition 
$e_{ij,\lambda}( \vec{k}) e_{ij,\lambda'}^*( \vec{k}) = \delta_{\lambda\lambda'}$. The second-order production from the scalar perturbations then gives, in the radiation dominated era \cite{Espinosa:2018eve},~\footnote{We note an additional factor $1/2$ in this solution with respect to the expression in \cite{Espinosa:2018eve}, which comes from the different normalization of the metric perturbation, which we adopt to be consistent with the notation of \cite{Maggiore:1999vm}.} 
\begin{equation} 
h_\lambda ( \eta ,\, \vec{k} ) =\frac{1}{2} \frac{4}{9 k^3 \eta} \int \frac{d^3 p}{\left( 2 \pi \right)^3}  
\, {\rm e}_\lambda^* ( \vec{k} ,\,\vec{p} ) \zeta ( \vec{p} )  \zeta ( \vec{k} - \vec{p} ) 
\left[ {\cal I}_c ( \vec{k} ,\, \vec{p} )  \cos \left( k \eta \right) 
+  {\cal I}_s ( \vec{k} ,\, \vec{p} ) 
\sin \left( k \eta \right) \right],
\label{h-sourced}
\end{equation} 
where ${\rm e}_\lambda ( \vec{k} ,\,\vec{p} ) \equiv {\rm e}_{ij,\lambda} ({\hat k} ) \vec{p}_i \vec{p}_j$, and where the two functions ${\cal I}_{c,s}$ have been computed analytically in \cite{Espinosa:2018eve,Kohri:2018awv}
\begin{align}
\label{eq: Ic, Is tau0=0}
\Ic(x,y) &= -36\pi\frac{(s^2+d^2-2)^2}{(s^2-d^2)^3}\theta(s-1)\ ,\\
\Is(x,y) &= -36\frac{(s^2+d^2-2)}{(s^2-d^2)^2}\left[\frac{(s^2+d^2-2)}{(s^2-d^2)}\log\frac{(1-d^2)}{|s^2-1|}+2\right], 
\end{align}
with 
\begin{equation}
d \equiv \frac{1}{\sqrt{3}}|x-y|, \qquad  s \equiv \frac{1}{\sqrt{3}}(x+y) ,\qquad  (d,s) \in [0,1/\sqrt{3}]\times[1/\sqrt{3},+\infty).
\label{eq: xy to ds}
\end{equation}
We insert these expressions into the GW energy density 
\cite{Maggiore:1999vm} 
\begin{equation}
\rho_\GW = \frac{M_p^2}{4} \left\langle \dot{h}_{ab} \left( t ,\, \vec{x} \right)  \dot{h}_{ab} \left( t ,\, \vec{x} \right) \right\rangle_T,
\label{rho}
\end{equation} 
where the dots denote differentiation with respect to physical time, and we  obtain  the expression (\ref{rhox}) in the main text. The GW polarization operators enter in this expression through the combination 
\begin{equation}
T \left[ {\hat k}_1 ,\, {\hat k}_2 ,\, \vec{p}_1 ,\, \vec{p}_2 \right] \equiv \sum_{\lambda_1,\lambda_2} 
e_{ij,\lambda_1} ( {\hat k}_1 )  e_{ab,\lambda_1}^* (  {\hat k}_1 ) 
e_{ij,\lambda_2} ( {\hat k}_2 )  e_{cd,\lambda_2}^* (  {\hat k}_2 ) \vec{p}_{1a} \vec{p}_{1b} \vec{p}_{2c} \vec{p}_{2d}. 
\label{def-T}
\end{equation} 
Using the identity 
\begin{align} 
2 \sum_\lambda e_{ij,\lambda} ( {\hat k})  e_{ab,\lambda}^* ( {\hat k}) &=
\left( \delta_{ia} - {\hat k}_i {\hat k}_a \right) \left( \delta_{jb} - {\hat k}_j {\hat k}_b \right) + 
\left( \delta_{ib} - {\hat k}_i {\hat k}_b \right) \left( \delta_{ja} - {\hat k}_j {\hat k}_a \right) \nonumber\\ 
&- \left( \delta_{ij} - {\hat k}_i {\hat k}_j \right) \left( \delta_{ab} - {\hat k}_a {\hat k}_b \right), 
\end{align} 
we obtain, after some algebra, 
\begin{align}
&
 T \left[ {\hat k}_1 ,\, {\hat k}_2 ,\, \vec{p}_1 ,\, \vec{p}_2 \right]  = 
\left[  \vec{p}_1 \cdot \vec{p}_2 
- {\hat k}_1 \cdot \vec{p}_1 \;\; {\hat k}_1 \cdot   \vec{p}_2 - {\hat k}_2 \cdot \vec{p}_1 \;\; {\hat k}_2 \cdot  \vec{p}_2
+ {\hat k}_1 \cdot {\hat k}_2 \;\; {\hat k}_1 \cdot \vec{p}_1 \;\;  {\hat k}_2 \cdot   \vec{p}_2  \right]^2  \nonumber\\ 
&
- \frac{1}{2}  \left[ p_2^2 -  \left( {\hat k}_2 \cdot \vec{p}_2 \right)^2 \right]  
\left[ 
p_1^2  - \left( {\hat k}_1 \cdot \vec{p}_1 \right)^2  - \left( {\hat k}_2 \cdot \vec{p}_1 \right)^2 
+ 2 \, {\hat k}_1 \cdot {\hat k}_2 \;\; {\hat k}_1 \cdot \vec{p}_1 \;\; {\hat k}_2 \cdot \vec{p}_1 
- \left( {\hat k}_1 \cdot {\hat k}_2 \right)^2 \left( {\hat k}_1 \cdot \vec{p}_1 \right)^2 
\right]  \nonumber\\ 
&
- \frac{1}{2}  \left[ p_1^2 -  \left( {\hat k}_1 \cdot \vec{p}_1 \right)^2 \right] 
\left[ 
p_2^2   - \left( {\hat k}_1 \cdot \vec{p}_2 \right)^2  - \left( {\hat k}_2 \cdot \vec{p}_2 \right)^2 
+ 2 \, {\hat k}_1 \cdot {\hat k}_2 \;\; {\hat k}_1 \cdot \vec{p}_2 \;\; {\hat k}_2 \cdot \vec{p}_2 
- \left( {\hat k}_1 \cdot {\hat k}_2 \right)^2   \left( {\hat k}_2 \cdot \vec{p}_2 \right)^2 
\right]   \nonumber\\ 
&
+ \frac{1}{4}  \left[ 1 + \left( {\hat k}_1 \cdot {\hat k}_2 \right)^2 \right] 
\left[ p_1^2 - \left( {\hat k}_1 \cdot \vec{p}_1 \right)^2 \right] 
\left[ p_2^2 - \left( {\hat k}_2 \cdot \vec{p}_2 \right)^2 \right]. 
\label{res-T}
\end{align} 
%

\section{Connected contributions to GW energy density two-point function}
\label{app:connected}
\begin{figure}[tbp]
	\centering 
	\includegraphics[width=1 \textwidth]{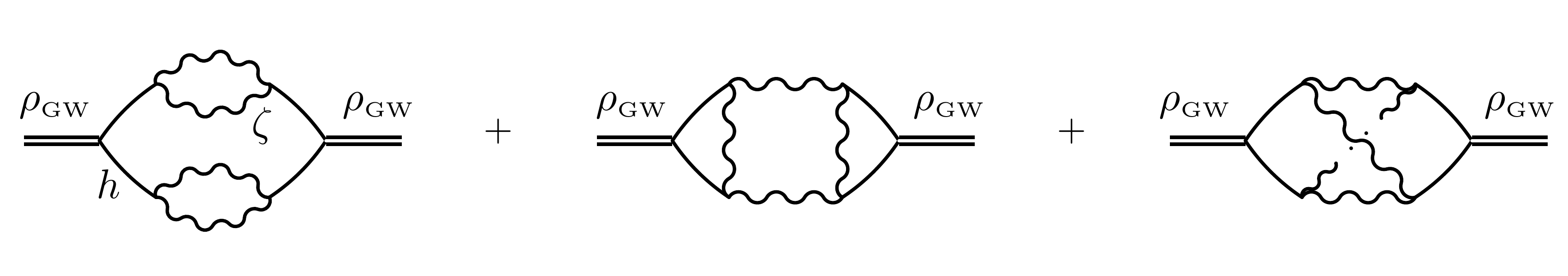}
	\caption{ Connected diagrams in the energy density two-point functions, case B, C, D respectively.
	}
	\label{fig:diagrams}
\end{figure}

In this appendix we give a sketch of the contribution of the connected diagrams of the energy density two-point function, giving rise to an anisotropy at extremely small scales.
We compute the two-point function starting from the definition of the energy density operator in Eq.~\eqref{rhox} as
\begin{align}
	&
	\langle \rho_\GW \left( \eta_1 ,\, \vec{x} \right)  \rho_\GW \left( \eta_2 ,\, \vec{y} \right) \rangle
	= 
	\lp \frac{ M_p^2}{81  a^2(\eta_1) a^2(\eta_2) \eta_1 \eta_2 }  \rp^2 \, \int \frac{d^3 k_1 d^3 k_2 d^3 p_1 d^3 p_2}{\left( 2 \pi \right)^{12} }
	\frac{1}{k_1^2 k_2^2} 
	\, {\rm e}^{i \vec{x} \cdot \left( \vec{k}_1 + \vec{k}_2 \right)} T \left[ {\hat k}_1 ,\, {\hat k}_2 ,\, \vec{p}_1 ,\, \vec{p}_2 \right] \nonumber\\ 
	& 
	\times\, \int \frac{d^3 k_3 d^3 k_4 d^3 p_3 d^3 p_4}{\left( 2 \pi \right)^{12} }
	\frac{1}{k_3^2 k_4^2} 
	\, {\rm e}^{i \vec{y} \cdot \left( \vec{k}_3 + \vec{k}_4 \right)} T \left[ {\hat k}_3 ,\, {\hat k}_4 ,\, \vec{p}_3 ,\, \vec{p}_4 \right] 
	\left \langle
	\, \zeta_{ \vec{p}_1 }  \zeta_{  \vec{k}_1 - \vec{p}_1 } 
	\zeta_{ \vec{p}_2 }  \zeta _{ \vec{k}_2 - \vec{p}_2 }
	\, \zeta_ { \vec{p}_3 }  \zeta_{ \vec{k}_3 - \vec{p}_3 }
	\zeta _{\vec{p}_4}  \zeta_{  \vec{k}_4 - \vec{p}_4 }
	\right \rangle 
	\nonumber\\ 
	&
	\times\left\langle 
	\left[ {\cal I}_s ( \vec{k}_1 ,\, \vec{p}_1 ) \cos \left( k_1 \eta_1 \right) 
	-  {\cal I}_c ( \vec{k}_1 ,\, \vec{p}_1 ) \sin \left( k_1 \eta_1 \right) \right] 
	\left[ {\cal I}_s ( \vec{k}_2 ,\, \vec{p}_2 ) \cos \left( k_2 \eta_1 \right) 
	-  {\cal I}_c ( \vec{k}_2 ,\, \vec{p}_2 ) \sin \left( k_2 \eta_1 \right) \right] 
	\right\rangle_T  \nonumber \\ 
	&
	\times\left\langle 
	\left[ {\cal I}_s ( \vec{k}_3 ,\, \vec{p}_3 ) \cos \left( k_3 \eta_2 \right) 
	-  {\cal I}_c ( \vec{k}_3 ,\, \vec{p}_3 ) \sin \left( k_3 \eta_2 \right) \right] 
	\left[ {\cal I}_s ( \vec{k}_4 ,\, \vec{p}_4 ) \cos \left( k_4 \eta_2 \right) 
	-  {\cal I}_c ( \vec{k}_4 ,\, \vec{p}_4 ) \sin \left( k_4 \eta_2 \right) \right] 
	\right\rangle_T  
\end{align} 
where we have introduced the notation $\zeta_{\vec q}\equiv \zeta (\vec q) $.
Now we can perform the stochastic average of the 8-point correlator, which can be expressed as
\begin{eqnarray} \label{wick}
	&& \left \langle
	\, \zeta_{ \vec{p}_1 }  \zeta_{  \vec{k}_1 - \vec{p}_1 } 
	\zeta_{ \vec{p}_2 }  \zeta _{ \vec{k}_2 - \vec{p}_2 }
	\, \zeta_ { \vec{p}_3 }  \zeta_{ \vec{k}_3 - \vec{p}_3 }
	\zeta _{\vec{p}_4}  \zeta_{  \vec{k}_4 - \vec{p}_4 }
	\right \rangle  \nonumber\\ 
	&&= 4 
	\left\langle \zeta_{\vec{p}_1} \zeta_{\vec{p}_2} \right\rangle
	\left\langle \zeta_{\vec{k}_1 - \vec{p}_1} \zeta_{\vec{k}_2 - \vec{p}_2} \right\rangle
	\left\langle \zeta_{\vec{p}_3} \zeta_{\vec{p}_4} \right\rangle
	\left\langle \zeta_{\vec{k}_3 - \vec{p}_3} \zeta_{\vec{k}_4 - \vec{p}_4} \right\rangle \nonumber\\ 
	&& \;\; + 8  
	\left\langle \zeta_{\vec{p}_1} \zeta_{\vec{p}_3} \right\rangle
	\left\langle \zeta_{\vec{k}_1 - \vec{p}_1} \zeta_{\vec{k}_3 - \vec{p}_3} \right\rangle
	\left\langle \zeta_{\vec{p}_2} \zeta_{\vec{p}_4} \right\rangle
	\left\langle \zeta_{\vec{k}_2 - \vec{p}_2} \zeta_{\vec{k}_4 - \vec{p}_4} \right\rangle \nonumber\\ 
	&&  \;\; + 32 
	\left\langle \zeta_{\vec{p}_1} \zeta_{\vec{p}_2} \right\rangle
	\left\langle \zeta_{\vec{p}_3} \zeta_{\vec{p}_4} \right\rangle 
	\left\langle \zeta_{\vec{k}_1 - \vec{p}_1} \zeta_{\vec{k}_3 - \vec{p}_3} \right\rangle
	\left\langle \zeta_{\vec{k}_2 - \vec{p}_2} \zeta_{\vec{k}_4 - \vec{p}_4} \right\rangle \nonumber\\ 
	&& \; + 16  
	\left\langle \zeta_{\vec{p}_1} \zeta_{\vec{p}_3} \right\rangle
	\left\langle \zeta_{\vec{k}_1 - \vec{p}_1} \zeta_{\vec{k}_4 - \vec{p}_4}  \right\rangle  
	\left\langle \zeta_{\vec{k}_2 - \vec{p}_2} \zeta_{\vec{k}_3 - \vec{p}_3}  \right\rangle
	\left\langle \zeta_{\vec{p}_2} \zeta_{\vec{p}_4} \right\rangle
\end{eqnarray} 
where the first line indicates the disconnected contribution (case A), while the remaining lines correspond to the connected pieces (case B, C, D, respectively) and are plotted diagrammatically in Fig.~\ref{fig:diagrams}.

The case A gives rise to a constant contribution to the equal-time two-point function proportional to the square of the expectation value of the energy density field
as 
\begin{equation}
	\langle \rho_\GW \left( \eta ,\, \vec{x} \right)  \rho_\GW \left( \eta ,\, \vec{y} \right) \rangle_{A}
	=  \la \rho_\GW \left( \eta ,\, \vec{x} \right) \ra \la  \rho_\GW \left( \eta ,\, \vec{y} \right) \ra
	= \la \rho_\GW \left( \eta \right) \ra ^2.
\end{equation}
In order to show that the remaining diagrams are suppressed at large scales, in the following we start by computing diagram B.
Inserting the correlators of the curvature perturbation in terms of its power spectrum and performing the time averages, one finds the equal time two-point function
\begin{align}
	&
	\langle \rho_\GW \left( \eta ,\, \vec{x} \right)  \rho_\GW \left( \eta ,\, \vec{y} \right) \rangle_{B}
	= 
	\frac{1}{2^7 \pi ^4}
	\lp \frac{ M_p^2}{81   a^2 \eta^2}   \rp^2 
	\, \int d^3 k_1  d^3 p_1  d^3 k_4  d^3 p_4
	\, {\rm e}^{i \lp \vec{x} -\vec y \rp  \cdot \left(  \vec{k}_1- \vec{k}_4  \right)} 
	\nonumber\\ 
	& 
	\times\,
	\frac{1}{k_1^4 k_4^4} 
	\frac{1}{p_1^3} \frac{1}{p_4^3} \frac{1}{\left\vert\vec k _1- \vec p_1 \right \vert^3} \frac{1}{\left\vert\vec k _4- \vec p_4\right \vert^3}
	T \left[ {\hat k}_1 ,\, -\hat k_4 ,\, \vec{p}_1 ,\, -\vec{p}_4 \right] 
	T \left[ - \hat k_1 ,\, {\hat k}_4 ,\, -\vec{p}_1 ,\, \vec{p}_4 \right] 
	\nonumber\\ 
	&  
	\times  
	\Pz(p_1) \Pz(p_4) \Pz\lp\left\vert\vec k _1- \vec p_1\right \vert \rp \Pz\lp\left\vert\vec k _4- \vec p_4\right \vert \rp
	\nonumber\\ 
	&
	\times \left(  {\cal I}_{s1}  {\cal I}_{s2} +  {\cal I}_{c1}  {\cal I}_{c2} \right) 
	\left(  {\cal I}_{s3}  {\cal I}_{s4} +  {\cal I}_{c3}  {\cal I}_{c4} \right) 
	\frac{\sin \left( \Delta_{34} T \right)}{ T \Delta_{34}}
	\frac{\sin \left( \Delta_{12} T \right)}{ T \Delta_{12}},
\end{align}
where ${\cal I}_{c,s,i} = {\cal I}_{c,s}(\vec k_1, \vec p_1)$, $\Delta_{ij} = k_i - k_j$, and the $T-$dependent terms are the leading ones after performing the time average.
Now we explicitly insert a Dirac delta power spectrum, change the integration variables into 
$\vec q=  \vec k_1 -\vec k_4$ and $ \vec s=\vec p_1-\vec k_1$ and rotate the reference frame such that $\vec q$ is aligned with the $\hat z$ axis, obtaining
\begin{align}
	&
	\langle \rho_\GW \left( \eta ,\, \vec{x} \right)  \rho_\GW \left( \eta ,\, \vec{y} \right) \rangle_{B}
	= 
	\frac{1}{4 (2 \pi)^3 }
	\lp \frac{ M_p^2 A_s^2}{81   a^2 \eta^2}   \rp^2 
	\int  d q q^2   j_0(q \, k_* \, \left \vert \vec x - \vec y \right \vert)
	\int d \Omega_{s} \int d \Omega_{p_1} \int d \Omega_{p_4}
	\nonumber\\ 
	&
	\times \delta\lp\left\vert  q \, {\hat e}_z  + {\hat  s}- {\hat p}_1+ {\hat  p}_4 \right \vert - 1 \rp
	\,
	\frac{1}{\left \vert {\hat p}_1 - {\hat s} \right \vert^4 } 
	\frac{1}{\left \vert q \, {\hat e}_z + {\hat s} - {\hat p}_1\right \vert^4 } 
	T^2 \left[ \frac{ {\hat p}_1 - {\hat s}}{\left \vert {\hat p}_1 - {\hat s} \right \vert} ,\, 
	\frac{ q \, {\hat e}_z + {\hat s} - {\hat p}_1}{ \left \vert  q \, {\hat e}_z + {\hat s} - {\hat p}_1 \right \vert} ,\, {\hat p}_1 ,\, -{\hat p}_4 \right] 
	\nonumber \\
	&
	\times   \Bigg[
	{\cal I}_{s} \lp \frac{1}{\left \vert {\hat p}_1 - {\hat s} \right \vert},
	\frac{1}{\left \vert {\hat p}_1 - {\hat s} \right \vert} \rp {\cal I}_{s} 
	\lp \frac{1}{\left \vert q \, {\hat e}_z + {\hat s} - {\hat p}_1\right \vert},\,
	\frac{1}{\left \vert q \, {\hat e}_z + {\hat s} - {\hat p}_1\right \vert} \rp 
	+ \nonumber\\ 
	& {\cal I}_{c} \lp \frac{1}{\left \vert {\hat p}_1 - {\hat s} \right \vert},
	\frac{1}{\left \vert {\hat p}_1 - {\hat s} \right \vert} \rp {\cal I}_{c} 
	\lp \frac{1}{\left \vert q \, {\hat e}_z + {\hat s} - {\hat p}_1\right \vert},\,
	\frac{1}{\left \vert q \, {\hat e}_z + {\hat s} - {\hat p}_1\right \vert} \rp 
	\Bigg]^2
	\left[ \frac{\sin \left( \Delta_{12} \, T \right)}{ \Delta_{12} \, T} \right]^2
\end{align} 
with 
\begin{equation} 
	\Delta_{12} \, T =  
	\Delta_{34} \, T = \left( k_1 - k_4 \right) T = 
	\left\{  
	\left\vert {\hat p}_1 - {\hat s} \right\vert - 
	\left\vert q \, {\hat e}_z + {\hat s} - {\hat p}_1 \right\vert 
	\right\} \, k_* \, T,
\end{equation} 
where we have redefined $\vec{q} = k_* \vec{q} \,'$ and  dropped the prime.

The spherical Bessel function plays a role of a window function that forces its argument to be of order one, and so 
\begin{equation} 
	q \sim \frac{1}{k_* \left\vert \vec{x}-\vec{y} \right\vert} \ll 1 
\end{equation}
since we are looking at anisotropies on scales $\left\vert \vec{x}-\vec{y} \right\vert \gg \frac{1}{k_*}$.
In the limit of small external momentum $q$, the time averaged term goes to 1 and we have
\begin{eqnarray}
	\langle \rho_\GW \left( \eta ,\, \vec{x} \right)  \rho_\GW \left( \eta ,\, \vec{y} \right) \rangle_{B}
	\simeq 
	\frac{1}{4 (2 \pi)^3 }
	\lp \frac{ M_p^2 A_s^2}{81   a^2 \eta^2}   \rp^2 \times 
	\pi \left(  \frac{1}{k_* \left\vert \vec{x}-\vec{y} \right\vert} \right)^3
	\times  {\cal S}_{B}
\end{eqnarray} 
where we defined 
\begin{align}
	{\cal S}_{B}&=
	\int d \Omega_{s} \int d \Omega_{p_1} \int d \Omega_{p_4}
	\delta\lp\left\vert  {\hat  s}- {\hat p}_1+ {\hat  p}_4 \right \vert - 1 \rp \frac{1}{\left \vert {\hat p}_1 - {\hat s} \right \vert^8 } 
	\nonumber\\ 
	& \times \,
	T^2 \left[ \frac{ {\hat p}_1 - {\hat s}}{\left \vert {\hat p}_1 - {\hat s} \right \vert} ,\, 
	-\frac{ {\hat p}_1 - {\hat s}}{ \left \vert  {\hat p}_1 -  {\hat s}  \right \vert} ,\, {\hat p}_1 ,\, -{\hat p}_4 \right] 
	\Bigg[
	{\cal I}^2 \lp \frac{1}{\left \vert {\hat p}_1 - {\hat s} \right \vert},
	\frac{1}{\left \vert {\hat p}_1 - {\hat s} \right \vert} \rp 
	\Bigg]^2.
\end{align}
 The integral can be performed numerically, and expressing the results in terms of the GWs density contrast, one finds finally
\begin{eqnarray}
	\langle \delta_\GW \left( \eta ,\, \vec{x} \right)  \delta_\GW \left( \eta ,\, \vec{y} \right) \rangle_{B}
	\simeq 
	2 \cdot 10^2
	\left(  \frac{1}{k_* \left\vert \vec{x}-\vec{y} \right\vert} \right)^3
\end{eqnarray} 
which is highly suppressed. A similar suppression is expected for the other two connected diagrams, which therefore give a negligible contribution to the GWs anisotropy.

\end{document}